\documentstyle[12pt,epsf]{article}
\newcommand{\tresj}[6]{\left( \begin{array}{ccc}
                              #1 & #2 & #3 \\
                              #4 & #5 & #6 
                             \end{array}
                      \right)}
\newcommand{\seisj}[6]{\left\{ \begin{array}{ccc}
                              #1 & #2 & #3 \\
                              #4 & #5 & #6 
                             \end{array}
                      \right\} }

\newcommand{\nk}{{\bf k}}
\newcommand{\np}{{\bf p}}
\newcommand{\nq}{{\bf q}}
\newcommand{\nr}{{\bf r}}

\begin{document}
\begin{titlepage}
\thispagestyle{empty}
\mbox{} 
\vspace*{2.5\fill} 

{\Large\bf 
\begin{center}

Effects of Short-Range Correlations in $(e,e'p)$ 
reactions and nuclear overlap functions

\end{center}
} 

\vspace{1\fill} 

\begin{center}
{\large M. Mazziotta, J. E. Amaro and F. Arias de Saavedra}
\end{center}

{\small 
\begin{center}
 {\em Departamento de F\'{\i}sica Moderna, 
          Universidad de Granada,
          Granada 18071, Spain}  \\   
\end{center}
} 

\kern 1.5cm 
\hrule 
\kern 3mm 

{\small\noindent
{\bf Abstract} 
\vspace{3mm} 

A study of the effects of short-range correlations over the $(e,e'p)$
reaction for low missing energy in closed shell nuclei is presented.
We use correlated, quasi-hole overlap functions extracted from the
asymptotic behavior of the one-body density matrix, containing central
correlations of Jastrow type, up to first-order in a cluster
expansion, and computed in the very high asymptotic region, up to 100
fm.  The method to extract the overlap functions is checked in a
simple shell model, where the exact results are known.  We find that
the single-particle wave functions of the valence shells are shifted
to the right due to the short-range repulsion by the nuclear core. The
corresponding spectroscopic factors are reduced only a few percent
with respect to the shell model.  However, the $(e,e'p)$ response
functions and cross sections are enhanced in the region of the maximum
of the missing momentum distribution due to short-range correlations.

} 

\kern 3mm 
\hrule 


\end{titlepage}

\newpage

\setcounter{page}{1}


\section{Introduction}


In the last years, quasi-free $(e,e'p)$ reactions have proved to
be an ideal tool to study the spectral function of nuclei
\cite{Kel96,Bof96}.  Apart from the ambiguities coming from
final-state interactions and off-shell effects in the electromagnetic
current, it is possible to extract such valuable information about
single-particle properties as momentum distributions and spectroscopic
factors of the different nucleon shells near the Fermi level
\cite{Van91}.

The observed values of the spectroscopic factors for closed shell
nuclei are around $S\sim 0.6$--0.7 \cite{Lap93}--\cite{Leu94} ---yet
theoretical studies appear to indicate that these values increase due
to relativistic effects \cite{Udi93}. The small values of the
spectroscopic factors are attributed to the departure from the extreme
single-particle model, where the nucleon is ejected from a well
defined orbit within the target nucleus.

Recent Variational Monte Carlo calculations including short-range
correlations in the nuclear wave function report spectroscopic factors
differing from unity only a few percent \cite{Rad94}--\cite{Mut00}.
In addition, center of mass correlations can enhance the spectroscopic
factor by a factor around 7\% for the valence shell of $^{16}$O
\cite{Van97,Die74}.  Although it has been shown that these values
could be lowered by the inclusion of low-energy configuration mixing
in the wave functions \cite{Geu96}, the experimental spectroscopic
factors, and hence the cross section of $(e,e'p)$ reactions for low
missing energy, are not satisfactorily explained by present
theoretical models.

It has been shown recently \cite{Van93} that the overlap functions
between the nuclear ground state with $A$ nucleons and a residual
state with $A-1$ nucleons ---which are a main ingredient in nucleon
knock-out calculations, including implicitly the spectroscopic
factor--- can be obtained from the asymptotic behavior of the
one-body density matrix (OBDM). This was confirmed in an exactly
solvable many-body system in one dimension with a zero-range
interaction \cite{Van96}.  The theorem was applied in \cite{Sto96} to
obtain the overlap functions and spectroscopic factors from a model
OBDM including Jastrow correlations, and later \cite{Dim97,Gai00}
applied to the calculation of cross sections for several reactions,
such as $(p,d)$, $(e,e'p)$ and $(\gamma,p)$.  However, some problems
with the numerical restoration procedure performed in \cite{Sto96},
which is based on an exponential fit of the OBDM, were pointed out in
\cite{Van97}, namely: i) the use of harmonic-oscillator single-particle
wave functions, which do not have the correct exponential behavior in
the asymptotic region, to construct the OBDM, and ii) the nature of
the exponential fit allowed to obtain in \cite{Sto96} overlap
functions for the unoccupied states, while in \cite{Van97} it was
demonstrated that such extraction is not possible starting from a
CBF-type wave function.

In reference \cite{Gai00} the same restoration procedure was applied
to compute overlap functions using several correlated OBDM taken from
the literature.  In particular, the one obtained in \cite{Ari97} by a
truncated cluster expansion of the radial multipoles $\rho_l(r,r')$ of
the OBDM. Using the above density, spectroscopic factors corresponding
to the $p_{1/2}$ and $p_{3/2}$ shells of $^{16}$O were reported
\cite{Gai00} to be identical in this model and equal to 0.981.  The
equality of these values is probably due to the fact that the
restoration procedure in \cite{Gai00} started from a OBDM for $l=1$
which (i) was not separated in spin partners $j=1/2,3/2$, and that
(ii) was computed up to distances of $r,r'=11$ fm, which are not large
enough to separate asymptotically the two contributions $p_{1/2}$ and
$p_{3/2}$.  This fact, together with the problems pointed out above,
makes possible that some of the effects attributed in \cite{Gai00}
to short-range correlations, actually be a consequence of the
extraction procedure of the overlap functions.

One of the motivations for this work is to clarify this situation, in
particular to explore the possibility of separating the two spin-orbit
partners $j=l\pm 1/2$ starting from a $l$-multipole of the OBDM. Note
that, in the correlated model of \cite{Ari97}, the underlying Slater
determinant was built with Woods-Saxon single-particle wave functions
including spin-orbit splitting, so the model can deal with different
overlap functions, with different energies, corresponding to the
$p_{1/2}$ and $1p_{3/2}$ shells in $^{16}$O or $d_{3/2}$ and $d_{5/2}$
in $^{40}$Ca.

Another aim of this paper is to perform a numerical study of the
convergence of the asymptotic methods used to extract the overlap
functions from the OBDM. We first carry out such study in the nuclear
shell-model, where the exact overlap functions are known (they are
equal to the single-particle wave functions).  This allows us to
determine optimum upper values of the coordinates in which one should
know the OBDM in order to obtain convergence in the extraction
procedure.  We continue evaluating more realistic overlap functions
starting from the correlated OBDM of \cite{Ari97}, computed up to
$\sim$ 100 fm, so that we can check the convergence of the results. In
particular, we will be able to obtain precise values of the
spectroscopic factors for quasi-hole states.

Finally, in this work the overlap functions resulting from the above
 task will be inserted in a model of the $(e,e'p)$ reaction, in order
 to evaluate the effects of short-range correlations over nuclear
 response functions and cross sections.  We use the DWIA model of
 \cite{Ama98a,Ama99a} which includes a new expansion of the
 relativistic electromagnetic current in powers of the missing
 momentum, but {\em it is not} expanded in $q$ or $\omega$.  Combined
 with relativistic kinematics, this relativized model was shown in
 ref. \cite{Ama96a} to give the same results as the relativistic Fermi
 gas for the electromagnetic inclusive responses in nuclear matter.
 Moreover, the present model was compared in \cite{Udi99} with a fully
 relativistic DWIA model of the reaction for $|Q^2| = 0.8$
 (GeV$/c)^2$, giving a reasonable description of the $A_{TL}$
 asymmetries recently measured in $^{16}$O \cite{Gao00}.

This paper is organized as follows.  In sect. 2 we summarize the DWIA
model we use for coincidence electron scattering, and its relation
with the overlap functions, with some details of the multipole
analysis of response functions placed in the appendix.  In sect. 3 we
present the different asymptotic procedures to extract the overlap
functions from the exact OBDM, and the correlated model of the OBDM
including short-range correlations of the Jastrow type.  In sect.~4 we
present a study of the reliability and convergence of the asymptotic
methods in the shell model.  The reader is directed to sect.~5 for
discussion of the results obtained with the correlated model, where
the effects of short-range correlations over $(e,e'p)$ observables,
overlap functions, and spectroscopic factors are analyzed, with a
brief application of the model to $(\gamma,p)$ reactions.  Finally,
our conclusions are summarized in sect. 6.

\section{DWIA model of $(e,e'p)$ reactions}

\subsection{Cross section and response functions}

In this section we summarize those aspects of the $(e,e'p)$
reaction that are of relevance for this paper.  
We consider the process  shown in the Feynman diagram
of figure~1. Here, an incident
electron with four-momentum $K^\mu_e=(\varepsilon,\nk_e)$ interacts
with a target nucleus, exchanging a virtual
photon with four-momentum given by
$Q^\mu=(K_e-K'_e)^\mu=(\omega,\nq)$, with
$K'^\mu_e=(\varepsilon',\nk'_e)$ the scattered electron four-momentum.
The outgoing proton with four-momentum $P'^\mu=(E',\np')$ is detected
in coincidence with the scattered electron.

\begin{figure}
\setlength{\unitlength}{1.2mm}
\begin{center}
\begin{picture}(90,60)(0,0)
        \put(10,18){$K_e^\mu$}
        \put(10,38){$K'^\mu_e$}
        \put(78,18){$|\Phi^{(A)}_0\rangle$}
        \put(78,38){$|\Phi^{(A-1)}_{\alpha}\rangle$}
        \put(45,35){\makebox(0,0){$Q^\mu$}}
        \put(45,25){\makebox(0,0){$\gamma$}}
        \put(15,10){$e$}
        \put(15,50){$e'$}
        \put(10,10){\vector(1,1){10}}
        \put(20,20){\line(1,1){10}}
        \put(30,30){\vector(-1,1){10}}
        \put(20,40){\line(-1,1){10}}
        \multiput(29.5,29)(1.92,0){15}{$\sim$}
        \thicklines
       \put(80,10){\line(-1,1){18}}
        \put(81,11){\line(-1,1){17}}
        \put(82,12){\line(-1,1){17}}
        \put(70,18){\line(0,1){4}}
        \put(74,22){\line(-1,0){4}}
        \put(61,32){\vector(1,2){8}}
        \put(66,42){\line(1,2){6}}
        \put(62,32){\line(1,1){19}}
        \put(64,32){\line(1,1){18}}
        \put(65,31){\line(1,1){18}}
        \put(68,40){\line(1,0){4}}
        \put(72,36){\line(0,1){4}}
        \put(62,30){\oval(6,4)}
        \put(66,50){${\bf p'}$}
\end{picture}
\end{center}
\caption{\small One-photon exchange diagram for the $(e,e'p)$
reaction.}
\end{figure}
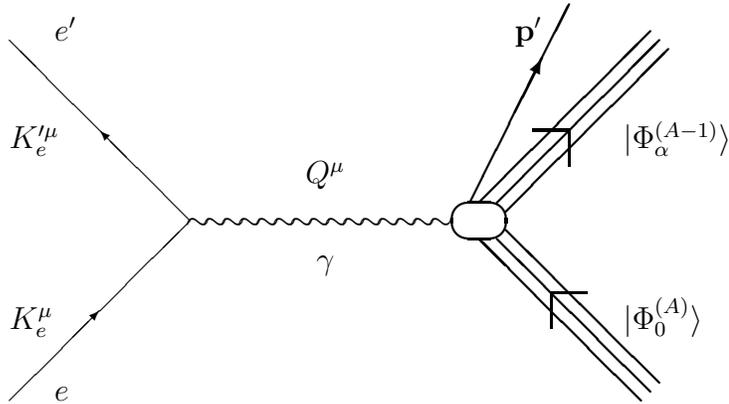

The wave function of the spin-zero target in the ground-state is
denoted $|\Phi^{(A)}_0\rangle$, with (non relativistic) energy
$E^{(A)}_0$.  We are interested in the low-missing energy region,
where the residual nucleus is left in a bound state,
$|\Phi^{(A-1)}_\alpha\rangle =|J_\alpha M_\alpha \rangle$, with (non
relativistic) energy $E^{(A-1)}_{\alpha}$.  We neglect recoil and
assume parity conservation.

We work in the laboratory system, with the $z$-axis pointing into the
$\nq$-direction and the $x$-axis in the electron scattering plane.  In
this reference system, the cross section for the $(e,e'p)$
reaction, assuming plane waves and the extreme relativistic limit for
the electron, can be written in the form \cite{Ras89}
\begin{equation}
\frac{d^5\sigma}{d\varepsilon' d\Omega'_e d\Omega_{p'}}
=\Sigma + h\Delta
\label{eq1}
\end{equation}
where $\Omega_{p'}=(\theta',\phi')$ are the proton emission angles,
$h$ is the electron helicity, 
$\Sigma$ is the unpolarized cross section and $\Delta$ is the electron 
polarization power. These functions are given by
\begin{eqnarray}
\Sigma
&=&
K\sigma_M\left[v_L W^L + v_T W^T + v_{TL} W^{TL}\cos\phi'
              +v_{TT} W^{TT}\cos2\phi'
         \right]
\\
\Delta
&=&
K\sigma_M  v_{TL'}W^{TL'}\sin\phi'.
\end{eqnarray}
Here the kinematical factor $K$ is proportional to the momentum $p'$,
\begin{equation}
K=\frac{p'M}{(2\pi\hbar)^3},
\end{equation}
$\sigma_M$ is the Mott cross section, 
and $v_K$ are  factors containing the dependence on the
electron kinematics:
\begin{equation}
\begin{array}{ll}
\displaystyle
v_L=\left({\frac{Q^{2}}{q^{2}}}\right)^2, &
\displaystyle
v_T=\tan^{2}\frac{\theta_e}{2}-\frac{Q^{2}}{2q^{2}}, \\
\displaystyle
v_{TL}=\frac{Q^2}{q^2}
       \sqrt{\tan^2\frac{\theta_e}{2}-\frac{Q^2}{q^2}}, &
\displaystyle
v_{TT}=\frac{Q^2}{2q^2}, \\
\displaystyle
v_{TL'}=\frac{Q^2}{q^2}\tan\frac{\theta_e}{2}. &
\end{array}
\end{equation}
Note that in this work the $v_{TL}$ and $v_{TL'}$ variables have an
extra $\sqrt{2}$ factor with respect to the corresponding definition
of ref. \cite{Ras89}.

The five exclusive nuclear response functions $W^K$ are defined by the
following linear combinations of longitudinal ($L$) and/or transverse
($T$) projections, with respect to the transfer momentum $\nq$, of the
relevant tensor describing the hadronic part of the emission mechanism
\begin{equation} 
\label{respuestas}
\begin{array}{l@{\kern 1cm}l}
\displaystyle
W^L = W^{00} & 
\displaystyle
W^T= W^{xx}+W^{yy} \\
\displaystyle
\cos\phi' W^{TL} = W^{0x}+W^{x0} &
\displaystyle
\cos 2\phi' W^{TT} = W^{yy}-W^{xx} \\
\displaystyle
\sin\phi' W^{TL'} = i(W^{0y}-W^{y0}) &
\end{array}
\end{equation}
Here the hadronic tensor, $W^{\mu\nu}$, is related to the transition matrix
elements of the nuclear electromagnetic 
current operator $\hat{J}^{\mu}(\nq)$.
It is defined by
\begin{equation}\label{hadronic-tensor}
W^{\mu\nu}=\frac{1}{K}
\sum_{m_s M_{\alpha}}
\langle \np' m_s,\Phi^{(A-1)}_{\alpha}|
\hat{J}^{\mu}(\nq)|\Phi^{(A)}_0\rangle
\langle \np' m_s,\Phi^{(A-1)}_{\alpha}|
\hat{J}^{\nu}(\nq)|\Phi^{(A)}_0\rangle
\end{equation}
and it represents the maximum information that can be obtained in
these kinds of experiments. Note that the dependence on the azimuthal angle of
the emitted proton, $\phi'$, is given 
explicitly in Eqs.~(\ref{respuestas}).
The final hadronic states entering in the definition of the hadronic
tensor, $| \np' m_s,\Phi^{(A-1)}_{\alpha}\rangle$ are in principle the
exact scattering states with the corresponding boundary conditions,
i.e., 
they correspond asymptotically  to a nucleon with momentum $\np'$ and 
third spin component $m_s$, and a daughter 
nucleus in the state $|\Phi^{(A-1)}_{\alpha}\rangle$.

\subsection{DWIA and overlap functions}

In this paper we consider a DWIA model of the current matrix elements
between hadronic states that enter into the hadronic tensor
(\ref{hadronic-tensor}). This model is based in the impulse
approximation, in which we assume that the nuclear electromagnetic
current is a one-body operator.  Hence we are neglecting two-body
contributions coming mainly from meson-exchange currents. The
contribution from these two-body currents was analyzed in
\cite{Ama99b}. However, in this work we are not including that
contribution, since we are interested in studying the genuine short-range
correlation effects.

The impulse approximation current operator
can be written in momentum space as
\begin{equation}\label{OB-current}
\hat{J}^{\mu}(\nq) 
= \int d^3k\, J^{\mu}(\nq+\nk,\nk)a^+_{\nq+\nk}a_{\nk}.
\end{equation}
where $J^{\mu}(\nq+\nk,\nk)$ is the single-nucleon current for which we use
a new non-relativistic expansion to first order in $\nk/M$, first
proposed in refs.  \cite{Ama96a,Ama96b}.  The time component of this
current contains charge and spin-orbit contributions, while the
transverse current is given as the sum of magnetization plus
convection pieces. However, each piece of the current differs from the
traditional non-relativistic one, containing, in addition to the
nucleon form factors, relativistic correction factors which depend on $q$
and $\omega$.  In ref. \cite{Ama96a} it was shown that,
beginning with the usual non-relativistic Fermi gas and using
relativistic kinematics plus the new currents, 
the same longitudinal and transverse inclusive response functions 
are obtained essentially as
in the relativistic Fermi gas model for arbitrary values of $q$,
which can be bigger than 1 GeV.  Note that the usual non-relativistic
forms of the current, that are also expanded in powers of $q/M$,
begin to fail for high values of $q>500$ MeV$/c$ and cannot be applied
for $q\sim M$.

The second approximation used in  DWIA is the description of the
ejected proton state as a single-particle wave function 
$\chi_{\np'}(\nr)$, obtained as the solution of the Schr\"odinger
equation with a complex  optical potential fitted to
elastic-scattering data. The final hadronic state is then written as
 \begin{equation}
|\Phi^{(A-1)}_{\alpha},\chi_{\np'}\rangle
=\int d^3 k \,\widetilde{\chi}_{\np'}(\nk) 
a^{\dagger}_{\nk}
|\Phi^{(A-1)}_{\alpha}\rangle
\end{equation}
where $\int d^3k\,\widetilde{\chi}_{\np'}(\nk) a^{\dagger}_{\nk}$ 
is the field operator creating a nucleon in the state
$\widetilde{\chi}_{\np'}(\nk)$, i.e.,  the wave function of the ejected
proton in momentum space.
From here we can write the matrix element of the current 
(\ref{OB-current}) as
\begin{equation}
\langle \Phi^{(A-1)}_{\alpha},\chi_{\np'}|
\hat{J}^{\mu}(\nq)
|\Phi^{(A)}_0\rangle 
= \int d^3 k \, d^3 k' \,
\widetilde{\chi}_{\np'}^*(\nk')
J^{\mu}(\nq+\nk,\nk)
\langle \Phi^{(A-1)}_{\alpha}|
a_{\nk'}
a^{\dagger}_{\nq+\nk}a_{\nk}
 |\Phi^{(A)}_0\rangle 
\end{equation}
Now we use the anti-commutation rules of the Fermion operators
\begin{equation}
a_{\nk'}a^{\dagger}_{\nq+\nk} = 
\delta(\nk'-\nq-\nk)
-a^{\dagger}_{\nq+\nk}a_{\nk'}.
\end{equation}
The contribution of the second term in this equation,
$a^{\dagger}_{\nq+\nk}a_{\nk'}$, can be neglected if

\begin{enumerate}

\item 
the momentum of the ejected proton is much larger than the Fermi
momentum of the initial nucleus, $p' \gg p_{F}$, this condition is
usually fulfilled in the experiments, and

\item the wave function $\widetilde{\chi}_{\np'}(\nk')$ of the ejected
nucleon is negligible outside of an interval of momentum $\Delta \nk'$
around the central value $\np'$.

\end{enumerate}

In such cases, we can write
\begin{equation}\label{orthogonality}
\widetilde{\chi}_{\np'}^*(\nk') a_{\nk'}|\Phi^{(A)}_0\rangle
\simeq 0 .
\end{equation}
This condition is equivalent to assume that the wave function
$\widetilde{\chi}_{\np'}$ is orthogonal to the components of the
initial state.  Non-orthogonality effects have been found to be small
in the region of low missing momentum. However, the assumptions done
in this approximation are not valid for high missing momentum, 
for which the region neglected by the
approximation (\ref{orthogonality}) is explored.

We are interested here in the low missing momentum region so 
we can write the current matrix element as
\begin{equation}\label{many-body-current}
\langle \Phi^{(A-1)}_{\alpha},\chi_{\np'}|
\hat{J}^{\mu}(\nq)
|\Phi^{(A)}_0\rangle 
= \int d^3 k \,
\widetilde{\chi}_{\np' }^*(\nk+\nq)
J^{\mu}(\nq+\nk,\nk)
\langle\Phi^{(A-1)}_{\alpha}|a_{\nk}|\Phi^{(A)}_0\rangle, 
\end{equation}
In this equation we identify the single-particle overlap function
between the states $\Phi^{(A)}_0$ and $\Phi^{(A-1)}_{\alpha}$,
defined, in momentum space, as the matrix element \cite{Kel96}
\begin{equation}\label{overlap-momento}
\widetilde{\Psi}_{\alpha}(\nk)
=  \langle\Phi^{(A-1)}_{\alpha}|a_{\nk}|\Phi_0^{(A)}\rangle. \ .
\end{equation}
Using this definition, we can write the many-body matrix element 
of the current (\ref{many-body-current})
as a matrix element between single-particle states, namely between 
the overlap function and the distorted wave of the final proton
\begin{equation} \label{corriente-overlap}
\langle \Phi^{(A-1)}_{\alpha},\chi_{\np'}|
\hat{J}^{\mu}(\nq)
|\Phi^{(A)}_0\rangle 
= \langle \chi_{\np'}|J^{\mu}(\nq)|\Psi_{\alpha}\rangle.
\end{equation}
This is the matrix element that we compute in the present work in
order to obtain the $(e,e'p)$ response functions.  The information
about short-range correlations is contained inside the overlap
functions $\Psi_{\alpha}$, which are obtained from 
a correlated OBDM
by the asymptotic method explained in the next section.  The matrix
elements (\ref{corriente-overlap}) are computed by performing a
multipole expansion of the current operators in terms of Coulomb,
electric and magnetic operators, and of the outgoing wave function
$\chi_{\np'}$ in partial waves.  The corresponding response functions
(\ref{respuestas}) are expanded in Legendre functions of $\cos\theta'$
(the angle between $\np'$ and $\nq$),
and their expressions are given in appendix A.

The physical interpretation of the overlap function is clear by
writing it in the form
\begin{equation}
\widetilde{\Psi}_{\alpha}(\nk) = \sqrt{S_{\alpha}} \phi_{\alpha}(\nk)
\end{equation}
where $\phi_{\alpha}$ is the overlap function normalized to the unity,
and it is usually identified with the effective single-particle wave
function of the ``shell'' occupied by the ejected nucleon.  The
spectroscopic factor $S_{\alpha}=\langle\Psi_{\alpha}|\Psi_{\alpha}\rangle$
is the occupancy probability of the shell.

The plane wave impulse approximation (PWIA)  will be useful for
the physical interpretation of the correlation effects shown below in
terms of overlap functions.  In PWIA since FSI is neglected, the wave
function of the ejected proton is a plane wave,
$\widetilde{\chi}_{\np'}(\nk) = \delta(\nk-\np')$, and hence
eq.~(\ref{many-body-current}) becomes
\begin{equation} 
\langle \Phi_{\alpha}^{(A-1)},\chi_{\np'}|
\hat{J}^{\mu}(\nq)
|\Phi_0^{(A)}\rangle 
=
J^{\mu}(\np',\np)
\widetilde{\Psi}_{\alpha}(\np),
\end{equation}
where we have introduced the missing momentum, $\np\equiv\np'-\nq$,
identified with the momentum of the proton before the interaction.  As
a consequence of the above factorization property, the hadronic tensor
(\ref{hadronic-tensor}) is proportional to the momentum distribution
$|\widetilde{\Psi}_{\alpha}(\np)|^2$ of the overlap function
\begin{equation}
W^{\mu\nu}=
w^{\mu\nu}(\np',\np)
\left|\widetilde{\Psi}_{\alpha}(\np)\right|^2,
\end{equation}
where $w^{\mu\nu}(\np',\np)$ is the hadronic tensor for a single
nucleon with initial momentum $\np$ and final momentum $\np'$:
\begin{equation}
w^{\mu\nu}(\np',\np)=
 J^{\mu}(\np',\np)^* J^{\nu}(\np',\np). 
\end{equation}
In the same way, the response functions are also proportional to the
momentum distribution:
\begin{eqnarray}
W^K 
& = & 
w^K(\np',\np) 
\left|\widetilde{\Psi}_{\alpha}(\np)\right|^2
\label{respuesta-factorizada}
\end{eqnarray}
where $w^K(\np',\np)$ are the response functions for electron
scattering by a single proton with momentum $\np$.

When the FSI is turned on, the above  factorization is not true anymore
but the general behavior of the response functions is preserved.
The mean effect of the FSI is a reduction of the responses due to the
absorptive part of the optical potential. We will see below that the effects
of the short-range correlations are decoupled from the FSI effects.

\section{Correlated model of OBDM and overlap functions}

\subsection{Overlap functions}

The basic quantities of interest for our calculations are the overlap
functions between nuclear states with $A$ and $A-1$ nucleons,
eq.~(\ref{overlap-momento}).  We work in coordinate space, where the
overlap function is
\begin{equation}\label{overlap-distruzione}
\Psi_{\alpha}(x) =
\langle\Phi^{(A-1)}_{\alpha}|a(x)|\Phi^{(A)}_0\rangle.
\end{equation}
Here $x=(\nr,s_z,t_z)$ is a generalized coordinate including spin and
isospin, $\nr$ is the relative coordinate respect to the center
of mass of the residual nucleus $\Phi^{(A-1)}_f$, and $a(x)$ is the
destruction operator of a nucleon at the point $x$. We assume that the
initial nucleus is in the ground state $\Phi^{(A)}_0$ with energy
$E^{(A)}_0$, while the residual nucleus remains in an arbitrary state
$\Phi^{(A-1)}_{\alpha}$, with energy $E^{(A-1)}_{\alpha}$.

Using the Schr\"odinger equation verified by the initial and final
nuclear states, a system of integro-differential equations for the 
overlap functions can be written \cite{Ber65}. However in the
procedure explained below to compute these functions for the
low-energy levels of the residual nucleus we only make use of its
asymptotic behavior, which is based on the following equation verified
by the overlap functions at large distances, $r\rightarrow\infty$
\begin{equation} \label{Schrodinger}  
-\frac{\hbar^2}{2\mu}\nabla^2\Psi_{\alpha}(\nr)+
(A-1)v(\nr)\Psi_{\alpha}(\nr) 
= [E^{(A)}_0-E^{(A-1)}_{\alpha}]\Psi_{\alpha}(\nr) 
\end{equation}  
where $v(\nr)$ is the $NN$ potential and 
$\nr=\nr_1-\nr_2$  is the relative coordinate. 
This equation means that the overlap function
behaves asymptotically as a single-particle interacting with the
$A-1$ nucleons of the residual system as if they were located at the
same position, namely at the center of mass of the residual
nucleus. Of course this is only valid for so large distances that it
is not possible to take notice of the small separation distances of
nucleons within the nucleus.

In the cases when the initial nucleus has spin zero and the parity
of nuclear states is a good quantum number, it is possible to separate
the overlap function in radial and spin-angular parts \cite{Ber65}
\begin{equation} 
\label{overlap-radial}
\Psi_{\alpha}(\nr) = \phi_{nlj}(r){\cal Y}_{ljm}(\theta,\phi).
\end{equation}
with $j=J_{\alpha}$ the spin of the final nuclear state
$\Phi^{(A-1)}_{\alpha}$, and $l=j\pm1/2$, depending on the parity of this
state.  The radial overlap function $\phi_{nlj}(r)$ (where the quantum
number $n$ is reminiscent of the later identification with shell-model
states) verifies for large distances a radial equation coming from
eq.~(\ref{Schrodinger}), with an asymptotic eigenvalue given as the
difference between the energies of the initial and residual nuclei
$E^{(A)}_0-E^{(A-1)}_{\alpha}$, which is a negative number for every
value of the excitation energy of the residual nucleus. Therefore the
overlap function behaves as a bound state and it has the
typical exponential decay
\begin{equation}
\phi_{nlj}(r) \sim C \frac{e^{-kr}}{r},
\kern 2cm
r\rightarrow\infty
\end{equation}
with
\begin{equation}  
k = 
\sqrt{\frac{2\mu \left|E^{(A-1)}_{\alpha}-E^{(A)}_0\right|}{\hbar^2}}.
\end{equation} 
The exponential decay is modified by a logarithmic phase in the case
of proton emission, where the Coulomb potential plays a role. However
this fact does not modify the following results.

The relation between the overlap functions and the OBDM follows from
the definition of the density matrix of the initial nucleus:
\begin{equation}
\rho(\nr_1,\nr_2)=
\sum_s \langle \Phi_0^{(A)}|
a^{\dagger}(\nr_1,s)a(\nr_2,s)
|\Phi_0^{(A)}\rangle
\end{equation}
Inserting a complete set of states $|\Phi_{\alpha}^{(A-1)}\rangle$ 
of the  residual nucleus between the two Fermi
operators we obtain an
expansion of the OBDM in terms of overlap functions
\begin{equation} \label{densidad-overlap}
\rho(\nr_1,\nr_2)=
\sum_{\alpha}
\Psi_{\alpha}^{\dagger}(\nr_1)
\Psi_{\alpha}(\nr_2)
\end{equation}
In this work we consider the OBDM expanded in multipole densities
with angular momentum $l$
\begin{equation}
\rho(\nr_1,\nr_2) = \frac{1}{4\pi}\sum_l
\rho_l(r_1,r_2)P_l(\cos\theta_{12})
\end{equation}
where $\theta_{12}$ is the angle between $\nr_1$ and $\nr_2$.
Inserting the expression (\ref{overlap-radial}) into
(\ref{densidad-overlap}) and performing the sums over third components
we find the corresponding expansion of the OBDM multipoles in terms of
radial overlap functions
\begin{equation} \label{densidad-overlap-l}
\rho_l(r_1,r_2)
=\sum_{nj}(2j+1)\phi_{nlj}(r_1)\phi_{nlj}(r_2)
=\sum_{nj}\psi_{nlj}(r_1)\psi_{nlj}(r_2)
\end{equation}
where we have defined $\psi_{nlj}(r)$ as the radial part of the 
overlap function normalized with a factor $\sqrt{2j+1}$. 
\begin{equation} \label{overlap-j}
\psi_{nlj}(r)= \sqrt{2j+1}\phi_{nlj}(r).
\end{equation}

\subsection{Asymptotic methods for computing the overlap functions}

We recall here the method presented in \cite{Van93} to compute the
overlap functions by means of the exact OBDM for the ground state of the 
$A$-particle system.

We consider a fixed value of $l$, and denote the corresponding radial
overlap functions with angular momentum $l$ simply as 
$\psi_{\alpha}(r)$, 
$\alpha=0,1,2,\ldots$, corresponding to bound states of the 
residual nucleus ordered by increasing energy
$E_0^{(A-1)} < E_1^{(A-1)} < \cdots$, and with asymptotic behavior
\begin{equation}\label{overlap-asintotica}
\psi_{\alpha}(a) \sim C_{\alpha}\frac{e^{-k_{\alpha}a}}{a},
\kern 2cm
a\rightarrow\infty
\end{equation}
with $k_{\alpha}=[2\mu(E_{\alpha}^{(A-1)}-E^{(A)}_0)]^{1/2}/\hbar$.

The asymptotic behavior of the OBDM is, from
eq.~(\ref{densidad-overlap-l}),
\begin{equation}
\rho_l(r,a) \sim 
\sum_{\alpha}\psi_{\alpha}(r)C_{\alpha}
\frac{e^{-k_{\alpha}a}}{a},
\kern 2cm
a \rightarrow\infty .
\end{equation}
Now, due to the ordering $k_0< k_1 < \cdots$, the above sum is
dominated by the first term, with the slowest exponential decay, for
long distances $a$ such that $a(k_1-k_0) \gg 1$
\begin{equation}\label{densidad-dominante}
\rho_l(r,a) \sim 
\psi_0(r)C_0
\frac{e^{-k_0a}}{a},
\kern 2cm
a \rightarrow\infty
\end{equation}
Combining this equation  with the asymptotic
behavior of the diagonal part, which allows us to determine the
constant $C_0$, 
\begin{equation}\label{densidad-diagonal}
\rho_l(a,a) \sim 
|C_0|^2
\frac{e^{-2k_0a}}{a^2},
\kern 2cm
a \rightarrow\infty,
\end{equation}
we can compute the overlap function $\psi_0(r)$ and the corresponding
separation energy.  In order to obtain the second overlap function, we
apply the procedure to the density obtained subtracting the
contribution of the first overlap
\begin{equation}
\rho_l(r,a)-\psi_0(r)\psi_0(a) 
\sim \psi_1(r) C_1 \frac{e^{-k_1a}}{a},
\kern 2cm a \rightarrow\infty .
\end{equation}
In principle all the overlap functions corresponding to bound states
of the residual nucleus may be obtained by repeating these steps. 
In the next section we check the validity of this procedure, which in
the present paper we call ``exponential decay method''.

There is an alternative, equivalent way \cite{Van97} of obtaining the
overlap functions from the asymptotic behavior of the OBDM, without
using explicitly the exponential decay property. We will illustrate it 
 using the fact
that the diagonal part of the OBDM has also an exponential decay given
by eq.~(\ref{densidad-diagonal}), from where we can write
\begin{equation}
\sqrt{\rho_l(a,a)} \sim 
|C_0|
\frac{e^{-k_0a}}{a},
\kern 2cm
a \rightarrow\infty,
\end{equation}
which is precisely the behavior of the first overlap
function, eq.~(\ref{overlap-asintotica}). 
Using this equation in the asymptotic form of the OBDM,  we obtain
\begin{equation}\label{sqrt-rho}
\psi_0(r) = \lim_{a\rightarrow\infty}
\frac{\rho_l(r,a)}{\sqrt{\rho_l(a,a)}}.
\end{equation}
Note that in the case $C_0<0$ we obtain a minus sign which is just a
global phase that can be inserted in the overlap function. 
The application of this expression, which we call ``$\sqrt{\rho}$-method'',
has clear advantages over the exponential decay one 
when it is used to compute the overlap functions from a model OBDM
without the correct asymptotic behavior 
(for instance, constructed with harmonic oscillator  
single-particle wave functions).

\subsection{Model of correlated OBDM}

In this work we compute the overlap functions of closed shell nuclei  
by applying the last method explained  to a correlated OBDM.
We use the  model of ref. \cite{Ari97} which includes short-range
correlations up to first order in a cluster expansion of the OBDM.
The density and momentum distributions  were compared with the 
FHNC calculation of ref. \cite{Ari96}, with a good agreement between 
both models.

We begin with the OBDM of the initial nucleus $|\Phi^{(A)}_0\rangle$,
written  as
\begin{equation}
\rho(x_1,x_2) = 
\frac{A}{\langle\Phi^{(A)}_0|\Phi^{(A)}_0\rangle}
\int{dx_2,...,dx_A}\Phi^{(A)}_0{}^*(x_1,...,x_A)\Phi^{(A)}_0(x_1,...,x_A)
\end{equation}
For the applications to $(e,e'p)$ reactions, we will only  
need the proton density, this is obtained inserting in 
the previous equation the projection operator 
$Q(1)= \frac{1}{2}(1+\tau_z(1))$.

Short-range correlations are introduced within the model
by the Jastrow ansatz for the nuclear wave function
\begin{equation}\label{funcion-onda-correlacionada}
\Phi^{(A)}_0(1,2,\ldots,A)=F(1,2,\ldots,A)\Phi_{Sl}^{(A)}(1,2,\ldots,A).
\end{equation}
Here $\Phi^{(A)}_{Sl}$ is a Slater determinant and $F$ is a
correlation function containing two-body central correlations
\begin{equation}\label{Jastrow}
F(1,...,A) = \prod_{i>j=1}^{A}f(r_{ij}),
\end{equation}
where $r_{ij}=|\nr_{i}-\nr_{j}|$, and the function 
$f(r_{ij})$ has a Gaussian functional dependence
\begin{equation}
\label{gaussiana}
f(r) = 1-A\exp(-Br^2)
\end{equation}
We use the parameters $A=0.7$ and $B=2.2$ fm$^{-2}$ which were fixed
in \cite{Ari96} by minimizing the nuclear binding energies for the
Afnan and Tang S3 interaction.  In
ref. \cite{Ari97} up to six spin-isospin correlation channels were
included. However, the numerical effort grows in the present case,
since we have computed the OBDM up to distances so large as 100 fm,
essential to separate the first overlap function in some cases.
This fact compelled us to reduce the number of correlation channels
and to use a Gaussian dependence, in order to perform analytically the
multipole expansion of the correlation function $f(r)$.

\begin{figure}[hptb]
\begin{center}
\leavevmode
\epsfbox[100 380 500 630]{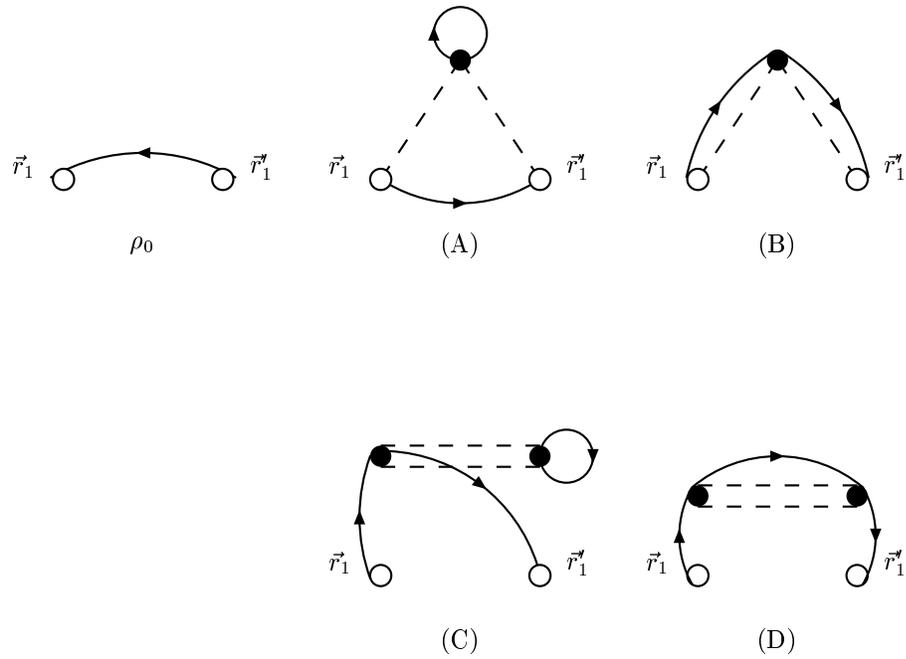}
\end{center}
\caption{\small
Diagrams considered in the cluster expansion of the OBDM. 
The dashed lines indicate the dynamical correlations $f(r_{ij})$
and the solid lines the uncorrelated density. }
\end{figure}

The OBDM is calculated by a cluster expansion, writing the 
correlation function as
\begin{equation}
F(1,\ldots,A)=1+\prod_{i>j}h(r_{ij})
\end{equation}
and performing an expansion up to second order in  $h$.
The resulting OBDM for protons can then be written as 
\begin{eqnarray}
\rho_1^p(\nr_1,\nr'_1)
&=&
\rho_0^p(\nr_1,\nr'_1)
+A(\nr_1,\nr'_1)
-B(\nr_1,\nr'_1)
-C(\nr_1,\nr'_1)
+D(\nr_1,\nr'_1)
\nonumber\\
&=&
\rho_0^p(\nr_1,\nr'_1)
+
\rho_0(\nr_1,\nr'_1)
\int d^3r_2
H(\nr_1,\nr'_1,\nr_2)
\rho_0(\nr_2,\nr_2)
\nonumber\\
&&
-\int d^3r_2
\rho_0(\nr_1,\nr_2)
H(\nr_1,\nr'_1,\nr_2)
\rho_0(\nr_2,\nr'_1)
\nonumber\\
&&
-\int d^3r_2
\int d^3r_3
\rho_0(\nr_1,\nr_2)
\rho_0(\nr_2,\nr'_1)
\rho_0(\nr_3,\nr_3)
H(\nr_2,\nr_2,\nr_3)
\nonumber\\
&&
+\int d^3r_2
\int d^3r_3
\rho_0(\nr_1,\nr_2)
\rho_0(\nr_2,\nr_3)
\rho_0(\nr_3,\nr'_1)
H(\nr_2,\nr_2,\nr_3)
\label{densidad-correlacionada}
\end{eqnarray}
where the new correlation function $H$ 
can be expressed in terms of the correlation function $f(r)$
\begin{equation}
\label{H}
H(\nr_i,\nr_j,\nr_k)=Q(1)[f(r_{ik})f(r_{jk})-1].
\end{equation}
This function contains the correlations between the two pairs
of particles $(ik)$ and  $(jk)$. The operator $Q(1)$  
guarantees that the particle 1 is a proton.

The functions $A,B,C,D$ corresponding to the corrections to the 
uncorrelated OBDM, $\rho^0(\nr_1,\nr'_1)$, 
are represented diagrammatically in Fig.~2. 
Therein, the open circles represent the two coordinates 
$\nr_1$ and  $\nr'_1$, while the solid dots refer to  
coordinates of inner  nucleons which are correlated to the rest.
A continuous line represents a non correlated OBDM
$\rho_0(\nr_1,\nr_2)$, while the dashed lines 
joins the different particles involved in the new correlation function $H$.
Thus in the diagrams $A$ and $B$, the particles 1 and 1'
are simultaneously correlated to a third particle 2.
On the other hand, in diagrams $C$ and $D$ there are two inner particles 2
and 3 which are correlated between them.

Using the above expression a multipole expansion is performed to
obtain the radial densities, $\rho_l(r,r')$, needed to compute the
overlap functions for different angular momenta.  We refer to \cite{Ari97}
for  details on this expansion. Note that, on the contrary to ref. 
\cite{Van97}, in this model we do not
separate the multipoles  $\rho_{lj}(r,r')$ 
of the density in spin-orbit partners $j=l\pm 1/2$ explicitly.
However these two
contributions are included in $\rho_l$. Since our correlated OBDM is
based in a single-particle basis obtained with a Woods-Saxon potential
including spin-orbit interaction, the two overlap functions
$\psi_{nlj}$
corresponding to an occupied shell  have different
energies from the beginning and, in principle, they can be separated 
in the asymptotic region.

\section{Test of the asymptotic methods in the shell model}

Before computing the overlap functions using the correlated model of
sect. 3.3, it is convenient to perform a test of the asymptotic
methods using a nuclear model where the exact solution is known {\em a
priori}.  In this way we will be able to determine (i) which of
the algorithms introduced below is more adequate to extract
the overlap functions, and (ii) the asymptotic distance needed 
to separate the several overlap functions. 

 We perform this analysis in the extreme shell model (SM), where the
overlap functions are just the single-particle wave functions of
the occupied shells.  Apart from its simplicity, another reason to 
choose the SM is that corresponds to the zero-order of the
correlated model.  Under the assumption that the Jastrow correlations,
as a first-order correction to the SM, do not drastically change the
asymptotic behavior of the OBDM, we expect that the convergence
conditions found in the SM will keep valid
in the correlated model.

\begin{table}[htbp]  
\begin{center}
\begin{tabular}{lllllll}
\hline\hline
Nucleus&     &$V_0$ [Mev]&$V_{ls}$ [Mev]&$R$ [fm]&$a_0$[fm]&$a_{ls}$ [fm]
\\\hline
       & $P$ & 62.00     & ~3.20        & 2.74   & 0.57    & 0.57\\[-1.5ex]
$^{12}$C \\[-1.5ex]
       & $N$ & 60.00     & ~3.15        & 2.74   & 0.57    & 0.57\\
\hline
       & $P$ & 52.50     & ~7.00        & 3.02   & 0.53    & 0.53\\[-1.5ex]
$^{16}$O \\[-1.5ex]
       & $N$ & 52.50     & ~6.54        & 3.02   & 0.53    & 0.53 \\
\hline
       & $P$ & 57.50     & 11.11        & 4.10   & 0.53    & 0.53 \\[-1.5ex]
$^{40}$Ca \\[-1.5ex]
       & $N$ & 55.00     & ~8.50        & 4.10   & 0.53    & 0.53 \\ 
\hline
       & $P$ & 59.50     & ~8.37        & 4.36   & 0.53    & 0.53 \\[-1.5ex]
$^{48}$Ca \\[-1.5ex]
       & $N$ & 50.00     & ~7.54        & 4.36   & 0.53    & 0.53 
\\\hline\hline
\end{tabular}
\end{center}
\caption {\small
Parameters of the Woods-Saxon potential.
} 
\end{table}

The single-particle wave functions in the SM are obtained by solving
the Schr\"odinger equation for a nucleon in a Woods-Saxon potential
\begin{equation}  
V_{WS}(r) 
= 
-\frac{V_0}{1+\exp(\frac{r-R}{a_0})}
-\vec{l} \cdot\vec{\sigma}
\left(\frac{\hbar}{m_{\pi}c}\right)^2
\frac{1}{r}\frac{d}{dr}
\left[\frac{V_{ls}}{1+\exp(\frac{r-R}{a_{ls}})}\right]
+V_C(r),
\end{equation}
where $V_C(r)$ is the Coulomb potential for protons, of an uniform
charge distribution with radius $R$.  The parameters of the potential
are given in table~1 for the closed shell nuclei
considered in this work.

We have solved numerically the radial equation up to distances of $r=100$
fm, in order to compute the corresponding OBDM in a wide asymptotic
region, where we will be able to check the convergence of the methods.
First, we compared our wave functions with the ones obtained integrating
the equation up to 11 fm, as is done traditionally, obtaining
essentially the same answer in both cases up to the region close to
$r\sim 11$ fm.  The energy eigenvalues of the proton shells are shown
in the third column of table~2.
For each value of $l$, we construct the shell model OBDM as a sum of
the corresponding single-particle radial densities of the occupied
states with angular momentum equal to $l$.  

\begin{table}[tb]
\begin{center}
\begin{tabular}{llr}\hline\hline
Nucleus        &  $nlj$   & $|E_{ws}|$ [Mev] \\\hline
$^{12}$C       &  $1s_{1/2}$ &  32.27   \\
               &  $1p_{3/2}$ &  15.49   \\\hline
$^{16}$O       &  $1s_{1/2}$ &  27.36   \\
               &  $1p_{3/2}$ &  13.92   \\
               &  $1p_{1/2}$ &   9.29   \\\hline
$^{40}$Ca      &  $1s_{1/2}$ &  35.54   \\
               &  $2s_{1/2}$ &   9.80   \\
               &  $1p_{3/2}$ &  26.12   \\
               &  $1p_{1/2}$ &  22.63   \\
               &  $1d_{5/2}$ &  15.83    \\
               &  $1d_{3/2}$ &   8.37    \\\hline
$^{48}$Ca      &  $1s_{1/2}$ &  39.25 \\
               &  $2s_{1/2}$ &  14.73  \\       
               &  $1p_{3/2}$ &  30.15 \\
               &  $1p_{1/2}$ &  28.00 \\
               &  $1d_{5/2}$ &  19.92 \\
               &  $1d_{3/2}$ &  15.14 \\\hline\hline
\end{tabular}
\end{center}
\caption{\small
Energies of the proton single particles in the shell model,
 $E_{WS}$.}
\end{table}

\subsection{Exponential decay methods.}

We first focus on the asymptotic decay method,  
in which  the overlap functions are obtained 
by fitting the exponential decay of the OBDM,
eq.~(\ref{densidad-dominante}). 
 This can be done in several ways:

\begin{description}
\item{(I) \bf Logarithm fit.}
Taking the logarithm in both sides of eqs. 
(\ref{densidad-dominante},\ref{densidad-diagonal}) we have asymptotically,
\begin{eqnarray}
\ln \{a|\rho_l(r,a)|\} &=& \ln \left\{C_0|\psi_0(r)|\right\} -k_0 a \\
\ln \{a^2\rho_l(a,a)\} &=& \ln C_0^2 -2k_0 a.
\end{eqnarray}
We first compute $C_0$ by fitting a straight line to $\ln\{a^2\rho_l(a,a)\}$
(we assume $C_0>0$ since this is just a
 global phase). Then $|\psi_0(r)|$ is computed by fitting another 
straight line
 to $\ln \{a|\rho_l(r,a)|\}$ and dividing by $C_0$. 
Finally, the sign of $\psi_0(r)$ is obtained 
 from  eq.~(\ref{densidad-dominante}) as the one of $\rho_l(r,a)$. 

\item{(II) \bf Trace minimization.} 
The overlap function is calculated  dividing the  OBDM 
by an exponential 
\begin{equation}
\psi_0(r) = \frac{\rho_l(r,a)}{C_0\frac{e^{-k_0a}}{a}}
\end{equation}
with $C_0$ and $k_0$ determined from the diagonal
part $\rho_l(a,a)$ as in  method I. The remaining parameter, 
$a$, is chosen by imposing that the overlap density 
$\rho_l^0(r,r')=\psi_0(r)\psi_0(r')$ 
be as close as possible to the OBDM, $\rho_l(r,r')$,
for every value of $r$ and $r'$ contained in an asymptotic
interval $[a_l,a_u]$. This condition is achieved by minimizing  the 
trace functional
\begin{equation} \label{traza}
F(a) \equiv {\rm Tr}[\rho_l-\rho_l^0]
= \int_0^{a_u} dr \, \int_{a_l}^{a_u} dr'\,
\left[ \rho_l(r,r')-
      \frac{\rho_l(r,a)\rho_l(r',a)}{C_0^2\frac{\exp(-2k_0a)}{a^2}}
\right]^2
\end{equation}

\item{(III) \bf Trace minimization with three parameters.} This third
method is a variation of fit II, that considers $C_0$ and $k_0$ 
as additional parameters in the trace functional. 
So the three parameters are now fixed by computing the absolute
minimum of this functional.
\end{description}

The fit procedure II is similar to the one applied in ref. \cite{Sto96}.
In the following we compare the results provided by these 
methods for different choices of the 
asymptotic interval $[a_l,a_u]$ where the fit is performed. 
We shall study the case of  $^{40}$Ca in the shell model, where the
OBDM contains $l=0,1,2$ multipoles.

\begin{figure}[tp]
\begin{center}
\leavevmode
\epsfbox[120 320 500 720]{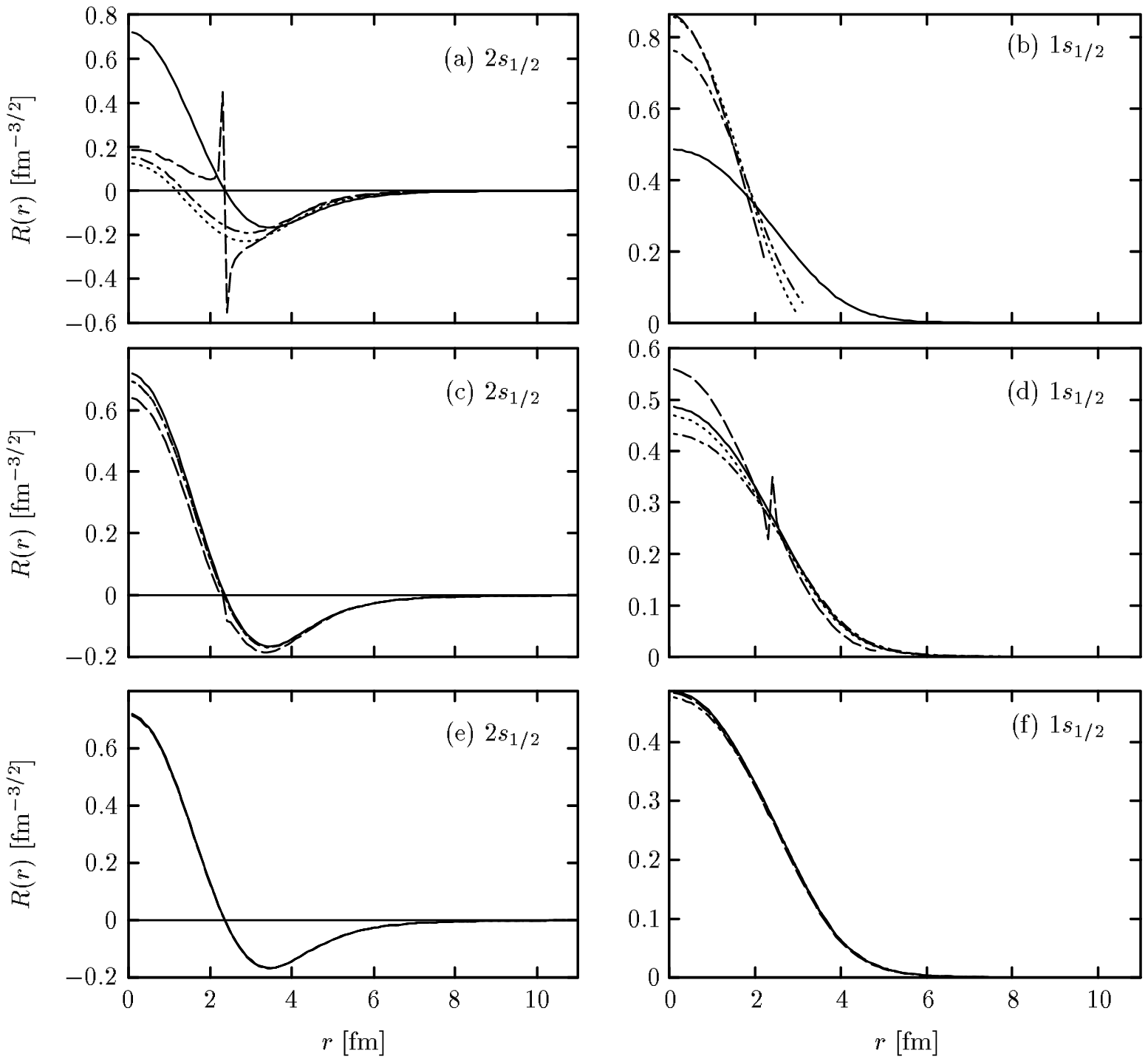}
\end{center}
\caption{\small
Overlap functions of $^{40}{Ca}$ for $l=0$ in shell model, computed
with the exponential decay method I (dashed lines), II (dot-dashed
lines) and III (dotted lines).
With solid lines we show the radial functions, corresponding to the
exact result. 
The several panels refer to different asymptotic intervals $[a_l,a_u]$  
considered in the fit of the first overlap function, namely
[2.3 fm, 11 fm] for  (a,b),
[5.9 fm, 15 fm] for  (c,d), and
[13.7 fm, 15 fm] for  (e,f).
}
\end{figure}

In Fig.~3 we show results for $l=0$, corresponding to the overlap
functions of the shells $2s_{1/2}$ (first overlap function) and
$1s_{1/2}$ (second overlap). Panel (a) shows an example of what is
obtained using a OBDM computed up to $a_u=11$ fm.  
In addition we have used $a_l=2.3$ fm,
which is the point where the density $\rho_0(r,r)$
reaches the 10\% of its maximum.  This value is not large enough to
be considered asymptotic  and the
resulting overlap functions are clearly incorrect.  In this figure we note
a misbehavior of fit I (dashed lines) in the region close to the node,
where eq.~(\ref{densidad-dominante}) is not valid, since the OBDM is
dominated there by the {\em second} overlap function.  
This misbehavior is not found in fits II (dot-dashed lines) and III
(dotted lines) because in both cases the exponential fit is
done globally and not point by point.
Since the first overlap has not been adequately extracted, we also obtain an
incorrect result for the second overlap, shown in panel (b).
We note in this panel that the displayed curves stop around 2--2.5 fm.   
The reason is that the subtracted diagonal part $\rho_0(r,r)-\phi_0(r)^2$
becomes negative in this region  as a consequence of the 
incorrect value of $\phi_0(r)$. 

The results for the first overlap function improve when we increase
$a_u$ to $15$ fm and $a_l$ to $5.9$ fm, corresponding this last distance
to the point where the density is 1\% of its maximum value,
as it is shown in panel (c).  
Even though there is a clear improvement with respect
to the results of panel (a), there is still a small difference with
respect to the exact result (solid line), which is larger for the
results of fit I.  These small differences are amplified when 
the second overlap function, shown in panel (d), is computed. Nevertheless
there is also a clear improvement respect to the former
results of (b).

In order to find a reasonable agreement with the
exact result we have to use $a_l=13.7$ fm, where the
density reaches the 10$^{-7}\%$ of its maximum value.  The
corresponding results are shown in panel (e), where we have used again
$a_u=15$ fm.  Although not seen in the scale of the
figure, the results of fits II and III are closer to the exact result
than the corresponding to fit I.  Finally, the second overlap function
is shown in panel (f), where we still note small differences with the
exact result specially for low $r$. The results for the second overlap
rely heavily on the adequacy of the computed first overlap function.
These small differences can be further minimized if a higher
value for the asymptotic points $a_l,a_u$ is utilized in computing the first
overlap. We will see this when we discuss the $\sqrt{\rho}$
method.

In the case of the second overlap for
$l=0$ we use a different interval $[a'_l,a'_u]$ from the one
considered in the corresponding first overlap.  The upper limit $a'_u$
is chosen as the point where the subtracted density
$\rho_0(r,r)-\psi_0(r)^2$ becomes negative, since this is a clear
indication that the first overlap is incorrect at this
point.  The lower limit $a'_l$ is chosen as the point where the
subtracted density is 10\% of its maximum value.  This value
is not critical in the cases in which the first overlap function is
incorrect, as there is no  improvement by changing
$a'_l$. On the other hand, in the cases in which we obtain a
reasonable result for the first overlap function, the result for the
second one is already quite good for low values of $a'_l$. This is a
consequence of the simplicity of the shell model where we are working
since in this case the subtracted density is factorizable as a product
of single-particle wave functions of the $1s_{1/2}$ shell
\begin{equation}
\rho_0(r,r')-\psi_0(r)\psi_0(r') = 2 R_{1s_{1/2}}(r)R_{1s_{1/2}}(r').
\end{equation}
and it is not necessary to separate a third
overlap function. Of course in the correlated case, where there are
extra contributions to the OBDM, one should be careful and choose a
value of $a'_l$ for which there is convergence.

\begin{figure}[htbp]
\begin{center}
\leavevmode
\epsfbox[120 320 500 720]{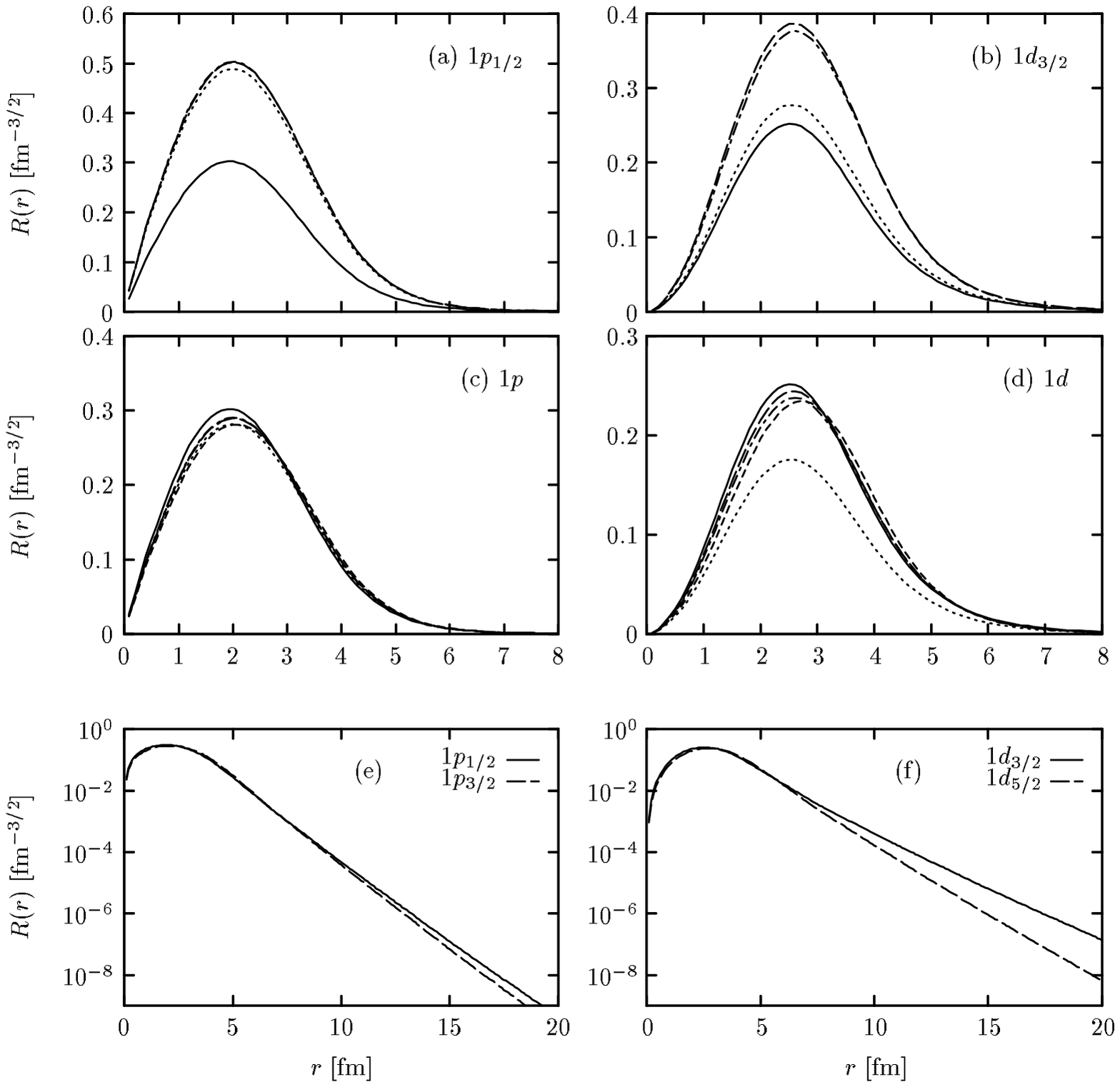}
\end{center}
\caption{\small 
First overlap functions of $^{40}$Ca for $l=1$ and
$l=2$ in shell model, computed with the exponential decay method II.
With solid lines we show the radial functions of the $1p_{1/2}$ and
$1d_{3/2}$ shells, corresponding to the exact results.  The asymptotic
intervals $[a_l,a_u]$ considered in the fit for $l=1$ are [4.0 fm, 11
fm] (dashed), [5.9 fm, 15 fm] (dot-dashed) and [13.0 fm, 20 fm]
(dotted); for $l=2$ they are [4.5 fm, 11 fm] (dashed), [6.6 fm, 15 fm]
(dot-dashed) and [17.2 fm, 20 fm] (dotted).  In panels (c) and (d) the
overlap functions have been ``averaged'' dividing by $[2(2l+1)]^{1/2}$
instead of $[2j+1]^{1/2}$, and also the wave functions of the
$1p_{3/2}$ and $1d_{5/2}$ shells are shown with short-dashed lines.
In panels (e) and (f) the asymptotic behavior of the $p$ and $d$ wave
functions has been displayed. 
}
\end{figure}

In Fig.~4 we show the first overlap functions of $^{40}$Ca for
$l=1,2$, corresponding to the shells $1p_{1/2}$ (a) and $1d_{3/2}$
(b).  The exact results are shown with solid lines, while with dashed,
dot-dashed and dotted lines we show respectively the results of the
fit II performed for three different asymptotic intervals with upper
limits $a_u=11, 15$ and 20 fm, and lower limits corresponding to the
points where the OBDM reaches the 10\%, 0.1\% and $10^{-9}$\% of its
maximum value.  In panel (a), corresponding to the $1p_{1/2}$ shell,
the three fits give a similar result which is around a factor of two
higher than the exact answer. The results do not show a noticeable
improvement when the asymptotic interval is increased from [5.9,15] fm
to [13,20] fm.  In panel (b), the results of fit II corresponding to 
the $1d_{3/2}$ shell are again above the exact
answer, although we note a convergence trend in going from the
interval $[6.6,15]$ (dot-dashed) to [17.2,20] fm (dotted).  The
results obtained with fits I and III are not shown in the figure; fit
III gives essentially the same result as fit II, while fit I is worse
than fit II.

It is clear from these results, that it is not possible to extract the
first overlap function for the $p$ and $d$ shells, using
asymptotic distances up to 20 fm.  This is related to the
fact that the energies of the spin-orbit partners ($1p_{1/2}$,
$1p_{3/2}$), and ($1d_{3/2}$, $1d_{5/2}$) have close values, their
difference being around 3.5 MeV and 7.5 MeV respectively 
(see table 2). Hence, at
20 fm the contribution of the second overlap function is still
important (see panels (e)--(f) in the same figure), specially in the
case of the $p$ shell where distances close to $100$ fm must be
used in order to get convergence, as we will discuss when we study the
$\sqrt{\rho}$ method.  Since in all these cases it was not possible to
extract reasonable values for the first overlap function, we do not
show the incorrect results for the second overlap function.

Recently Gaidorov {\em et al}.\ \cite{Gai00} have presented results
for the overlap functions of the $p$-shell in $^{16}$O using a method
similar to fit II.  The procedure was applied to the correlated OBDM
of \cite{Ari96} computed up to 11 fm, which uses the same shell model
we are considering here, with similar energies and wave functions for
the $1p_{1/2}$ and $1p_{3/2}$.  An averaged value for the $1p$ wave
function was extracted in that reference; we will explore this
possibility in the shell model.  In order to compute an averaged
overlap function we first assume that the multipoles of the OBDM
(\ref{densidad-overlap-l}) can be approximated by
\begin{equation}
\rho_l(r_1,r_2)
\simeq \sum_{nj}2(2l+1)\phi_{nl}(r_1)\phi_{nl}(r_2)
=\sum_{n}\psi_{nl}(r_1)\psi_{nl}(r_2)
\end{equation}
where $\phi_{nl}(r)$ is a mean value  of the two 
spin-orbit partners $\phi_{nlj}(r)$, with $j=l\pm 1/2$.  
The overlap function $\psi_{nl}(r)$ is now normalized with a factor 
$[2(2l+1)]^{1/2}$,  and we  assume  an asymptotic exponential behavior 
$\psi_{nl}(r) \sim C \exp(-kr)/r$. Hence we have for the OBDM
\begin{equation}
\rho_l(r,r')\sim \psi_{nl}(r)C\frac{e^{-kr'}}{r'},
\kern 2cm
r'\rightarrow \infty.
\end{equation}
Now we can proceed as before, by fitting an exponential decay 
to the OBDM,  and computing the averaged overlap function 
as $\phi_{nl}(r)= \psi_{nl}(r)/[2(2l+1)]^{1/2}$.

Results for the averaged overlap functions of the $1p$ and $1d$ shells
of $^{40}$Ca obtained by this procedure are shown in panels (c)--(d)
of Fig.~4.  The meaning of the lines and the asymptotic intervals used
in the fits are the same as in panels (a)--(b), but here we also
include, for comparison, the radial wave functions of the $1p_{3/2}$
and $1d_{5/2}$ shells with short-dashed lines. Note that the overlap
functions displayed in panels (c)--(d) are related to the ones of
(a)--(b) just by a global factor $[(2j+1)/2(2l+1)]^{1/2}$.

We begin discussing the results for the $1p$ shell shown in panel (c).
The fits shown with dashed and short-dashed lines are on average in
the intermediate region between the $1p_{1/2}$ and $1p_{3/2}$ curves.
However for higher values of the asymptotic interval (dotted lines)
the fit begin to move out of this region, ---it is now similar to the
$1p_{3/2}$ wave function--- and the possibility of obtaining an
average value breaks down.  The case of the $1d$ shell (d) is more
intriguing.  While dashed and short-dashed lines are between the
$d_{3/2}$ and $d_{5/2}$ curves, the dotted lines are well below
them.
The conclusion extracted from these results is that only for low
values of the asymptotic interval $[a_l,a_u]$ used in the fit a mean
value of the overlap function is provided.  However the results are
unstable, since they change when another interval in the fit is used,
and depend on the particular $l$-shell.

\begin{figure}[t]
\begin{center}
\leavevmode
\epsfbox[120 392 500 730]{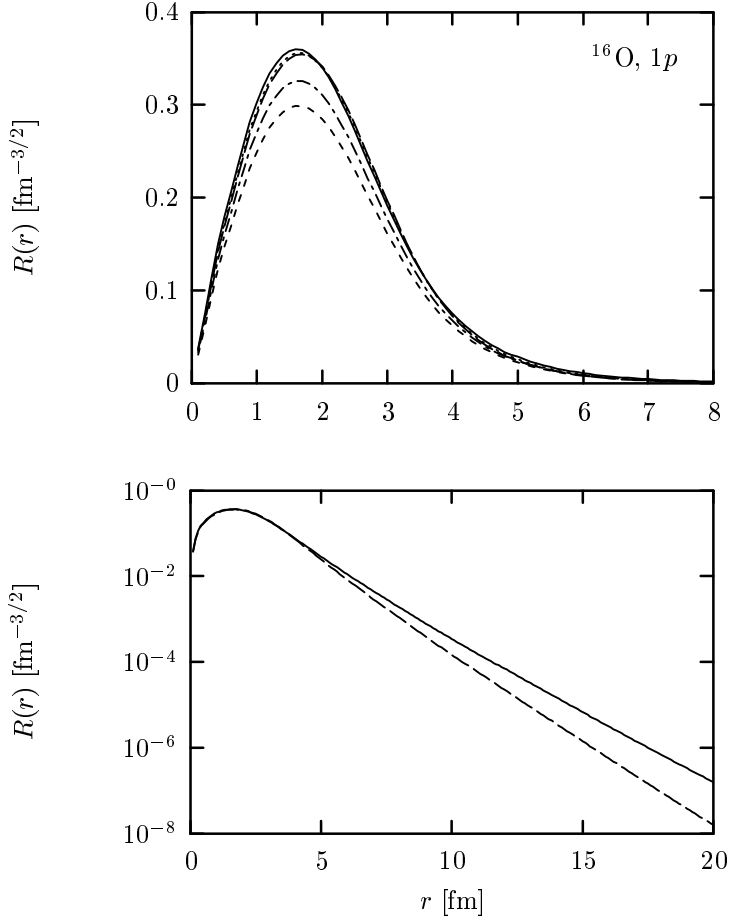}
\end{center}
\caption{\small Averaged overlap function of the $p$ shell of $^{16}$O
in shell model, computed using fit II in the asymptotic intervals
[3.5,11] fm (dotted lines), [9.5,15] fm (dot-dashed lines), and [12.2,
20] fm (short-dashed lines).  The exact functions of the $1p_{1/2}$
and $1p_{3/2}$ shells are shown with solid and long-dashed lines
respectively.}
\end{figure}

This behavior can be easily explained in the shell model.  For example
in the case of the $p$ shell, the exact OBDM is computed as
\begin{equation}
\rho_1(r,r') = 
 2R_{1p_{1/2}}(r)R_{1p_{1/2}}(r')
+4R_{1p_{3/2}}(r)R_{1p_{3/2}}(r).
\end{equation}
As seen in panel (e) of Fig. 4, the wave functions of the two partners
$1p_{1/2}$ and $1p_{3/2}$ are quite similar for low $r$. 
We note that they are also similar in the asymptotic region
up to $\sim 11$ fm, since their energies are very close. 
So we obtain the following approximation for the OBDM up to $\sim 11$
fm
\begin{equation}
\rho_1(r,a) \simeq 6 R_{1p}(r)R_{1p}(a)
\end{equation}
and for $a$ large, but not exceeding $\sim 11$ fm the two wave functions
have a similar exponential behavior  so we can write 
\begin{equation}\label{rho-media}
\rho_1(r,a) \sim \sqrt{6}R_{1p}(r) C\frac{e^{-ka}}{a},
\kern 1cm
a \rightarrow  11 fm
\end{equation}
and then it is possible to extract $R_{1p}(r)$ between the two
partners, as it is shown in panel (c) with dashed lines, where $a\leq 11$
fm.  For larger values of $a$ the two wave functions begin to separate
due to the different exponential decay.  For the interval [6.6, 15] fm
(dot-dashed lines in panel (c)) the two wave functions are still quite
close and eq.~(\ref{rho-media}) remain approximately valid. However,
for the dotted lines the asymptotic interval is [13,20] fm, where the
two wave functions are clearly different and eq.~(\ref{rho-media}) is
not valid. In this case the exact asymptotic behavior
\begin{eqnarray}
\rho_l(r,a) 
&\sim&  4R_{1p_{3/2}}(r)C_1\frac{e^{-k_1a}}{a}
      + 2R_{1p_{1/2}}(r)C_0\frac{e^{-k_0a}}{a}
\nonumber\\
&\simeq&  R_{1p}(r)\frac{4C_1e^{-k_1a}+2C_0e^{-k_0a}}{a}
\end{eqnarray}
should be used. The exponential fit of this equation may be performed, 
but the result will not be the searched quantity
$R_{1p}(r)$ and will depend on the interval $[a_l,a_u]$ used and on
the fitting method.  In
addition, in the very far limit where the second overlap can be
neglected, the fit procedure will converge to the exact first overlap
function $\psi_0(r)=
\sqrt{2j+1}R_{1p_{1/2}}(r)=\sqrt{2}R_{1p_{1/2}}(r)$.  Since the
pro-mediated overlap function is computed dividing to
$[2(2l+1)]^{1/2}=\sqrt{6}$,  by this procedure the
``averaged'' overlap function will converge to the wrong result
$\phi(r)= \sqrt{1/3}R_{1p_{1/2}}$.

Finally, in the $l=2$ case, the energy difference between the
$1d_{3/2}$ and $1d_{5/2}$ shells is bigger than in the case of the
$p$-shell. This makes that the two wave functions separate at shorter
distances $\sim 8$ fm, as seen in Fig. 4, panel (f).  This implies
that the results are less satisfactory than for the $p$-shell for high
asymptotic interval (see dot-dashed lines in panel (d)).  These are
clear indications of the impossibility of extracting an averaged
overlap function using this method.

The interesting example of  the $p$-shell in $^{16}$O is shown 
in Fig.~5.  Therein we show with solid and dashed lines
the corresponding wave functions of the $1p_{1/2}$ and $1p_{3/2}$ shells,
which are very close below 5 fm, where they begin to
separate. In the upper panel  we also show with
dotted lines the averaged overlap function obtained with fit II in the
asymptotic region [3.5,11] fm, which should correspond to the
mentioned calculation by Gaidorov {\em et al.,} in
ref. \cite{Gai00}. The dotted line is between the two $p$-shell wave 
functions, but if the same fit is applied to higher asymptotic intervals 
we obtain the dot-dashed and short-dashed curves of the figure, 
that are below the exact result. 

This example clarifies why the results of ref. \cite{Gai00} are
not far wrong in this particular case of $^{16}$O, since they used an
OBDM computed up to 11 fm. However the way in which the average is done 
is not under control. So in the correlated case one
should be careful in the interpretation of the results, because the
exact ones are not known {\em a priori}, and the effects due to
correlations cannot be unambiguously separated from the fit procedure.

\subsection{$\protect\sqrt{\protect\rho}$ method.}

Next we examine the alternative method based in eq.~(\ref{sqrt-rho}),
where the overlap function is directly computed as the quotient
\begin{equation}\label{sqrt-rho-method}
\psi_0(r) = \frac{\rho(r,a)}{\sqrt{\rho(a,a)}},
\end{equation}
for a value of $a$ large enough to reach convergence, which can be
easily checked by computing for several values of $a$.  
This method has several advantages over the exponential
fits previously analyzed. First it has not adjustable parameters and
no numerical minimization has to be carried out.  Second, the OBDM 
has not to be calculated in an interval $[a_l,a_u]$, but only in a few
asymptotic points $a$.  This is preferable in the correlated case,
where the computation of OBDM becomes longer.
Moreover, when the density is factorizable
$\rho(r,r')=\psi(r)\psi(r')$, then eq.~(\ref{sqrt-rho-method}) 
always provides the exact overlap function for any value of $a$.  
This makes this method exact in the shell model for all the shells 
in $^{12}$C and $^{16}$O.

\begin{figure}[hp]
\begin{center}
\leavevmode
\epsfbox[120 260 500 725]{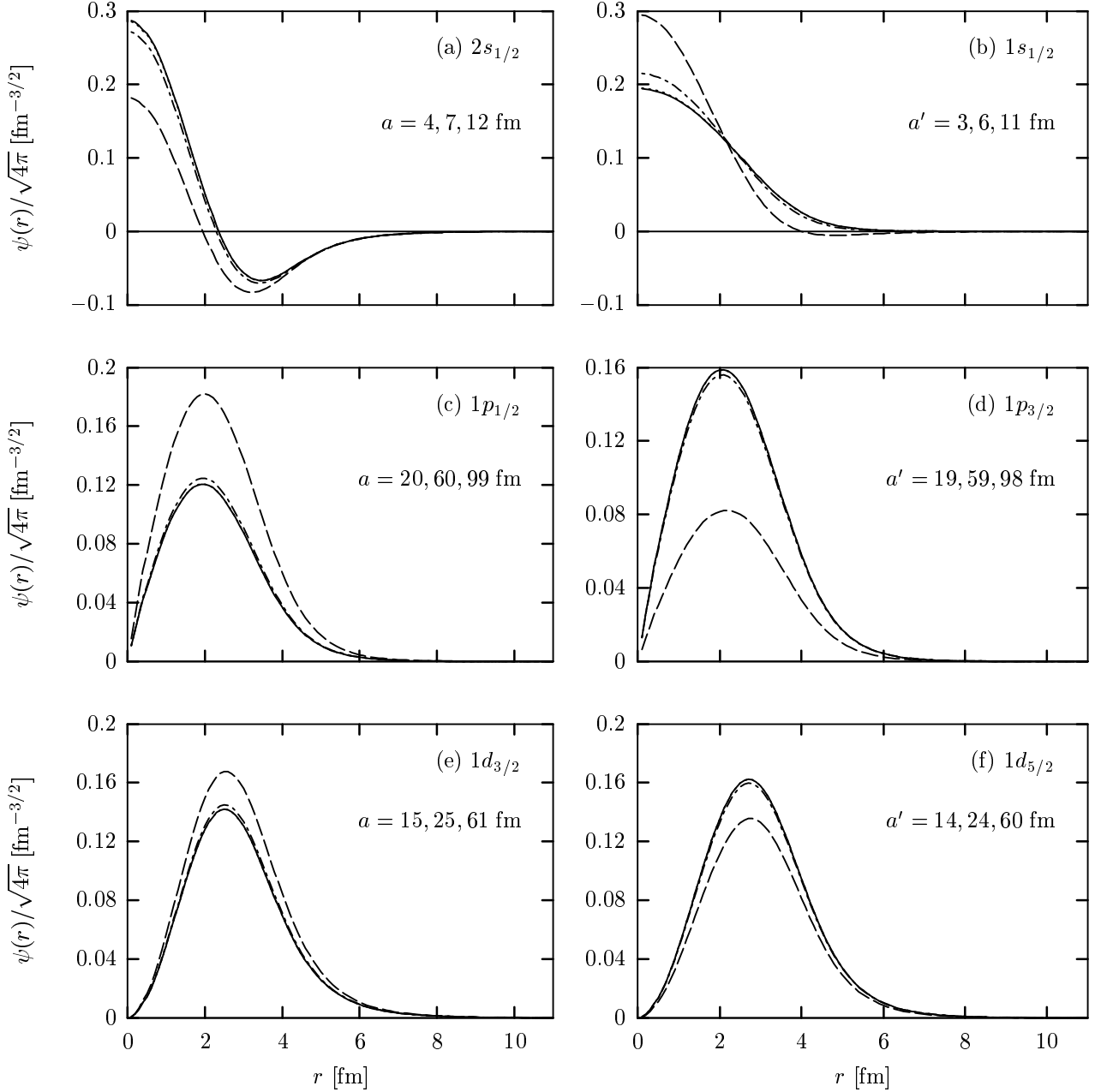}
\end{center}
\caption{\small Overlap functions of $^{40}$Ca in shell model computed
with the $\protect\sqrt{\protect\rho}$ method. All the functions are
normalized with a factor $\protect\sqrt{(2j+1)/4\pi}$. In each panel
we use the three indicated values of the asymptotic point $a$, that
correspond in ascending order to the dashed, dot-dashed and doted
lines respectively.  The exact result is shown with solid lines. The
second overlap functions (panels on the right) have been computed
using the first overlap function from the corresponding left panel and
a second asymptotic point $a'=a-1$ fm.  }
\end{figure}

In Fig.~6 we show results for the non trivial case of $^{40}$Ca in
which two overlap functions are present for each $l$. In this case the
asymptotic expression of the OBDM can be written 
\begin{equation}
\rho_l(r,a)\sim 
\frac{e^{-k_0a}}{a}
\left[
\psi_0(r)C_0
+\psi_0(r)C_1 e^{(k_0-k_1)a}
\right].
\end{equation}
and $a$ should be chosen large enough in order to 
neglect the second overlap function. An
appropriate value of $a$ can be
obtained by imposing $\exp[(k_0-k_1)a] =10^{-3}$. 
This makes the second overlap function contribution to be of the 
order $0.1\%$.  The estimated value of $a$ so obtained is
\begin{equation}\label{estimated}
a \simeq \frac{6.9}{k_1-k_0}.  
\end{equation}   
However, in practice the convergence will be reached at a different
value, depending on the relative value of the constants $C_0$ and
$C_1$. In the shell model we will determine the adequate value of $a$ by
comparing with the exact result.

The case of the $2s_{1/2}$ shell of $^{40}$Ca is shown in panel (a) of
Fig.~6. Therein we represent the first overlap function normalized
with a factor $\sqrt{(2j+1)/4\pi}$, and computed from
eq.~(\ref{sqrt-rho-method}) for three different values of $a$. Namely we
show the results for $a=4$ fm with dashed lines, $a=7$ fm with dot-dashed
lines, and $a=12$ fm with dotted lines. Using this last value
we already reproduce, within the scale of the figure, the exact
overlap function shown with solid lines.  This value is in agreement 
with $a\simeq 11.5$ fm provided by eq.~(\ref{estimated}). 

For each one of the curves presented in panel (a) we compute the
second overlap function using an asymptotic point $a'$ that must be
less than $a$. This is clear if we remember that the method matches
the asymptotic behaviors of the OBDM and of the first overlap
function for distances $r \ge a$. Then when we build the subtracted 
density  the asymptotic contribution will vanish in this region.
This is equivalent to say that the point $a$ effectively acts as 
the infinite point $a\simeq\infty$, so in this numerical method it 
has no sense to compute the second overlap function for $a'>a$.
For this reason, the second overlap functions for $l=0$ displayed in
panel (b) of Fig.~6, have been computed for asymptotic point $a'=a-1$
fm. Again for the biggest value shown, the extracted overlap function
almost coincides with the exact result shown with solid lines.

Results for the remaining shells $l=1,2$ are shown in panels (c)--(f).
As before, we show in each panel three curves corresponding to 
three ascending values of $a$  indicated in the figure, 
with dashed, dot-dashed and dotted lines respectively. In the case of
the second overlap function (panels on the right) we  use 
the corresponding first overlap function of the left panel and
asymptotic point $a'=a-1$ fm. The values of $a$ for which convergence
of the first overlap function is obtained 
 are shown in table~3.
In the case of the $p$-shell (panels (c) and (d)), we find convergence
for $a\sim 100$ fm.  Using the single-particle energies of table 2 and
eq.(\ref{estimated}), we obtain $a \simeq 86$ fm for the $p$- shell and 
$a\simeq 60$ for the $d$-shell. 

\begin{table}[ht]
\begin{center}
\begin{tabular}{lrrr}
\hline\hline
Nucleus    & $2s_{1/2}$ &  $1p_{1/2}$ & $1d_{3/2}$ 
\\\hline
$^{16}$O   &            &  80         &            
\\
$^{40}$Ca  & 12         &  100        & 61         
\\
$^{48}$Ca  & 12         & $> 100$     & 98         
\\\hline\hline
\end{tabular}
\end{center}
\caption {\small
Values of the asymptotic points for which convergence of the first
overlap function is reached with the $\protect\sqrt{\protect\rho}$
method.
} 
\end{table}

We have performed the same study of the proton overlap functions 
using the $\sqrt{\rho}$ method for
other closed-shell nuclei. 
The results of the convergence values for
 $^{16}$O and $^{48}$Ca  are summarized  in
table~3.
In the case of $^{16}$O we only show  the non-trivial case of
the $p$-shell. The corresponding overlap functions can be separated 
by using the OBDM computed up to $a\simeq 80$ fm. 
Finally, In the case of $^{48}$Ca, the separation energy within the $p$- and
$d$-shells
is smaller than in $^{40}$Ca (see table~2). As a consequence, the
convergence values are larger than in the former case. In particular,
for $a=100$ fm the $1p_{1/2}$ and $1p_{3/2}$ are not completely
separated and the convergence value is not shown in the table.
The estimated value of convergence for this shell is $a=138$ fm.

As a summary of this section,  with our present study using the shell
model, we have shown the reliability of the
asymptotic methods to compute the overlap functions of nuclei 
from the knowledge of the $l$-multipoles of the OBDM.
Our results have shown the
necessity of studying the convergence of the results in each case 
and that in many of them 
one should calculate the OBDM up to such huge distances as 100 fm in
order to separate the  overlap functions. 
In relation to the several methods studied, all of them
provide the correct result if the asymptotic interval $[a_l,a_u]$ 
is within the region of convergence. However, due to its simplicity, 
the $\sqrt{\rho}$ method is preferable in the 
general case in which the OBDM is the solution of a correlated
many-body problem, and this is the method we will use in the next
section to compute the correlated overlap functions.


\section{Results for $(e,e'p)$ observables and 
         overlap functions in the correlated case}


In this section we present results for overlap functions,
spectroscopic factors, $(e,e'p)$ response functions, and cross
sections, using the correlated model introduced in section 3.  Thus we
go beyond the single-particle model and will be able to identify the
effects of short-range correlations on these quantities and
observables.

\subsection{Quasi-hole overlap functions}

We start with the correlated OBDM of closed-shell nuclei, defined by
 eq.~(\ref{densidad-correlacionada}), and compute the multipoles,
 $\rho_l(r,r')$, as shown in ref.~\cite{Ari97}.  The zero-order
 density $\rho_l^0(r,r')$ in eq.~(\ref{densidad-correlacionada})
 corresponds to the non correlated shell model of sect.~4.  The
 correlated overlap functions for quasi-hole states can
 be obtained by using the $\sqrt{\rho}$ method discussed in the
 last section.  We apply eq. (\ref{sqrt-rho-method}) to the correlated
 OBDM for asymptotic points, $a$, large enough to reach convergence.
In the present case the exact result are not known {\em a priori},
but we are guided by the former study performed in the
 shell model. It is  expected that the values of the convergence asymptotic 
points in the correlated case do not change too much respect to
 the shell model ones. 
This can be understood in our model 
by studying the correlated
OBDM, eq.~(\ref{densidad-correlacionada}). 
Using the fact that the correlation function $f(r)\rightarrow 1$ for 
$r\rightarrow\infty$, we have for the function $H$
 defined in (\ref{H})
\begin{equation}
H(\nr_1,\nr'_1,\nr_2) 
\sim Q(1)[f(r_{12})-1],
\kern 1cm
r'_1\rightarrow\infty.
\end{equation}
On the other hand, the non-correlated density is dominated
asymptotically by the first overlap function of the shell model
\begin{equation}
\rho_0(\nr_1,\nr'_1) \sim \phi_0(\nr_1)\phi_0(\nr'_1)
\kern 1cm
r'_1\rightarrow\infty.
\end{equation}
From these equations it is straightforward to  
obtain the following asymptotic expression for
the correlated density (\ref{densidad-correlacionada})
for $r'_1\rightarrow \infty$
\begin{equation}
\rho_1^p(\nr_1,\nr'_1)
\sim 
K(\nr_1) \phi_0(\nr'_1) 
\kern 1cm
r'_1\rightarrow\infty
\end{equation}
where the function $K(\nr_1)$ is defined as
\begin{eqnarray}
K(\nr_1)
&\equiv &
\phi_0(\nr_1)
+
\phi_0(\nr_1)
\int d^3r_2
Q(1)[f(r_{12})-1]
\rho_0(\nr_2,\nr_2)
\nonumber\\
&&
-\int d^3r_2
\rho_0(\nr_1,\nr_2)
Q(1)[f(r_{12})-1]
\phi_0(\nr_2)
\nonumber\\
&&
-
\int d^3r_2
\int d^3r_3
\rho_0(\nr_1,\nr_2)
\phi_0(\nr_2)
\rho_0(\nr_3,\nr_3)
H(\nr_2,\nr_2,\nr_3)
\nonumber\\
&&
+
\int d^3r_2
\int d^3r_3
\rho_0(\nr_1,\nr_2)
\rho_0(\nr_2,\nr_3)
\phi_0(\nr_3)
H(\nr_2,\nr_2,\nr_3)
\label{funcion-K}
\end{eqnarray}
We can also determine the asymptotic behavior of this function $K(\nr_1)$
by using the particular Gaussian form (\ref{gaussiana}) 
of the correlation function, so for 
$r_1\rightarrow\infty$
the second and third terms in (\ref{funcion-K}) 
can be neglected with respect to
the other terms due to its Gaussian decay. Then we can write
\begin{equation}
K(\nr_1) \sim \eta \phi_0(\nr_1),
\kern 1cm
r_1\rightarrow\infty
\end{equation}
where the constant $\eta$ is defined as
\begin{eqnarray}
\eta
&\equiv &
1-
\int d^3r_2
\int d^3r_3
|\phi_0(\nr_2)|^2
\rho_0(\nr_3,\nr_3)
H(\nr_2,\nr_2,\nr_3)
\nonumber\\
&&
+
\int d^3r_2
\int d^3r_3
\phi_0(\nr_2)
\rho_0(\nr_2,\nr_3)
\phi_0(\nr_3)
H(\nr_2,\nr_2,\nr_3).
\end{eqnarray}
Using now the $\sqrt{\rho}$ method,  the first correlated overlap function
reads
\begin{equation}
\psi_0(\nr)=
\lim_{r'\rightarrow\infty}
 \frac{ K(\nr)\phi_0(\nr') }{ \sqrt{K(\nr')\phi_0(\nr')}}
= \frac{K(\nr)}{\sqrt{\eta}}.
\end{equation}
Since the function  $K(\nr)\sim \eta\phi_0(\nr)$, for 
$\nr\rightarrow\infty$,
we see that this correlated overlap function 
behaves asymptotically as the non-correlated one multiplied by the
constant $\sqrt{\eta}$
\begin{equation} \label{correlated-asymptotic}
\psi_0(\nr)\sim \sqrt{\eta}\phi_0(\nr), 
\kern 1cm
r\rightarrow\infty
\end{equation} 
Hence in the present model the short-range correlations do not modify
the energy of the first overlap function respect to the shell
model, since the asymptotic behavior of the OBDM is determined by the
exponential decay of the single-particle wave function $\phi_0(\nr)$.
Note that for shorter distances the above proportionality
(\ref{correlated-asymptotic}) does not hold because, in that case, the
function $K(\nr)$ includes other terms depending on the non correlated
density and on the correlation function $f(r)$, as can be seen in
eq. (\ref{funcion-K}).

The same procedure can be applied to each one of the multipoles 
$\rho_l(r,r')$ to show that the energy of the first overlap function
for each value of $l$ does not change respect to the uncorrelated
case. 
The same conclusion was also obtained in ref. \cite{Van97} 
in a model similar to ours by starting with  the OBDM $\rho_{lj}(r,r')$.
However in our model it is not possible to prove easily a
similar result for the energy of the {\em second} overlap function. 

These arguments indicate that the asymptotic points needed to compute
the overlap functions are similar to the ones found in the shell
model. In any case, in our calculations we have checked numerically
the convergence of the results for the different overlap functions.
For this reason, the correlated OBDM has been computed for values of
the asymptotic point as high as 100 fm in order to separate the first
overlap function in the cases in which the energies of two overlap
functions are close in the shell model. The values of the asymptotic 
point $a$ where convergence is reached are given in the third column of 
table 4. We first note that the convergence values of the
asymptotic point $a$ are similar to the ones obtained in the shell
model (compare with table 3).  Thereby, in order to extract the
$1p_{1/2}$ overlap function, we need to go up to $\sim 86$ fm for
$^{16}$O and up to $\sim 100$ fm for $^{40}$Ca, while in the case of
the $1d_{3/2}$ overlap function for $^{40}$Ca, convergence is found
for $a\sim 64$ fm.  This indicates that in fact the separation
energies of correlated overlap functions are close to the ones of the
shell model.

\begin{figure}[t]
\begin{center}
\leavevmode
\epsfbox[120 590 500 725]{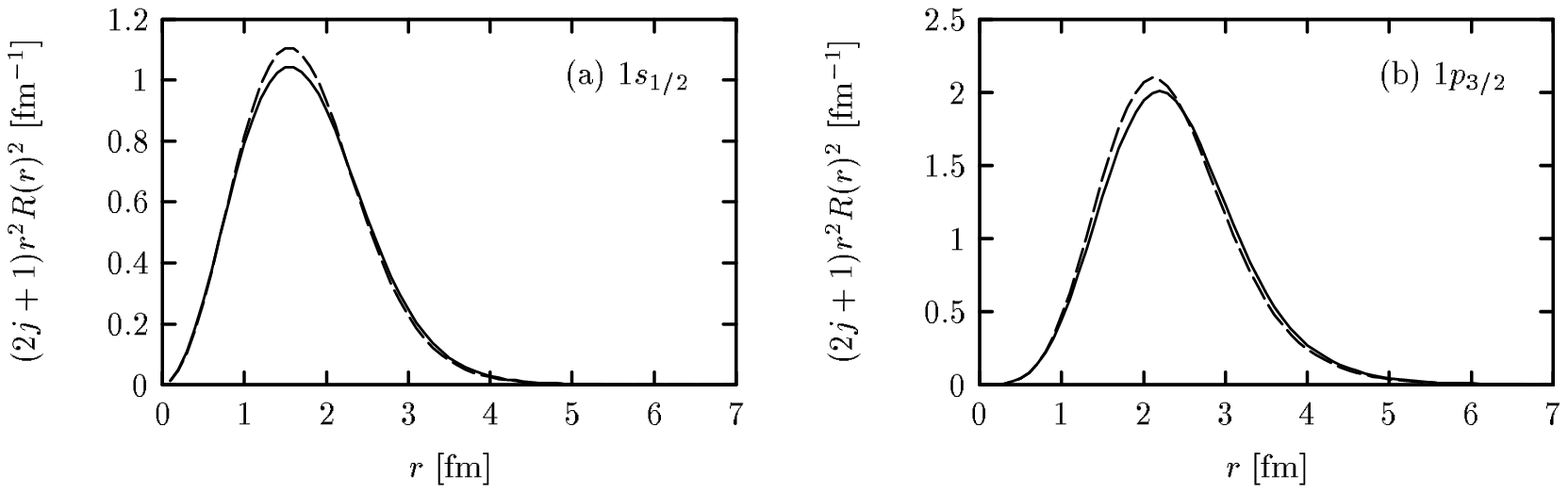}
\end{center}
\caption{\small  Solid lines: radial density of overlap
functions for $^{12}$C computed from the correlated OBDM.
With dashed lines we show the non-correlated results in the shell model.
}
\end{figure}

\begin{figure}[t]
\begin{center}
\leavevmode
\epsfbox[120 420 500 725]{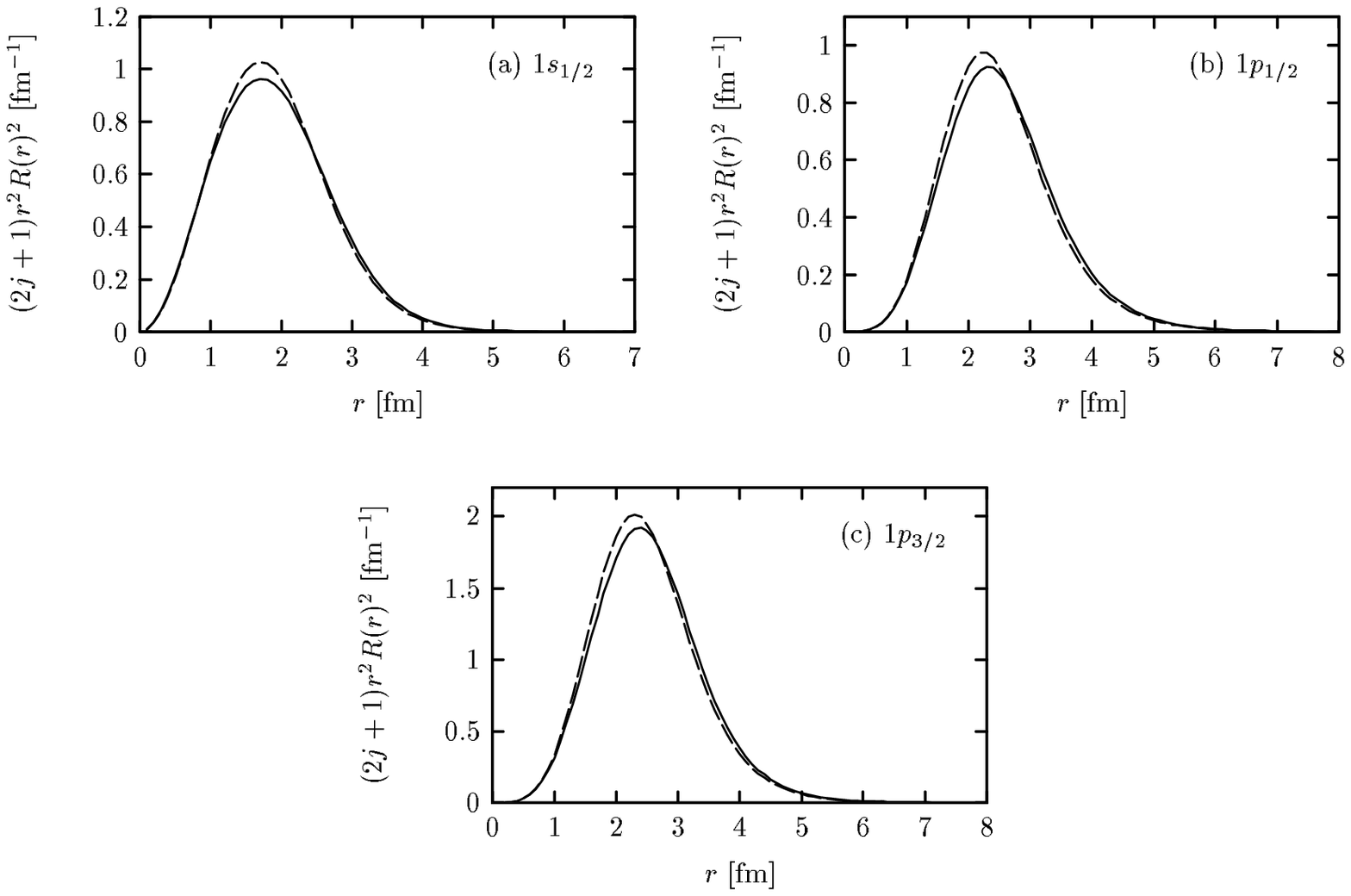}
\end{center}
\caption{\small 
The same as fig. 7 for the nucleus $^{16}$O.
}
\end{figure}

\begin{figure}[hp]
\begin{center}
\leavevmode
\epsfbox[120 260 500 725]{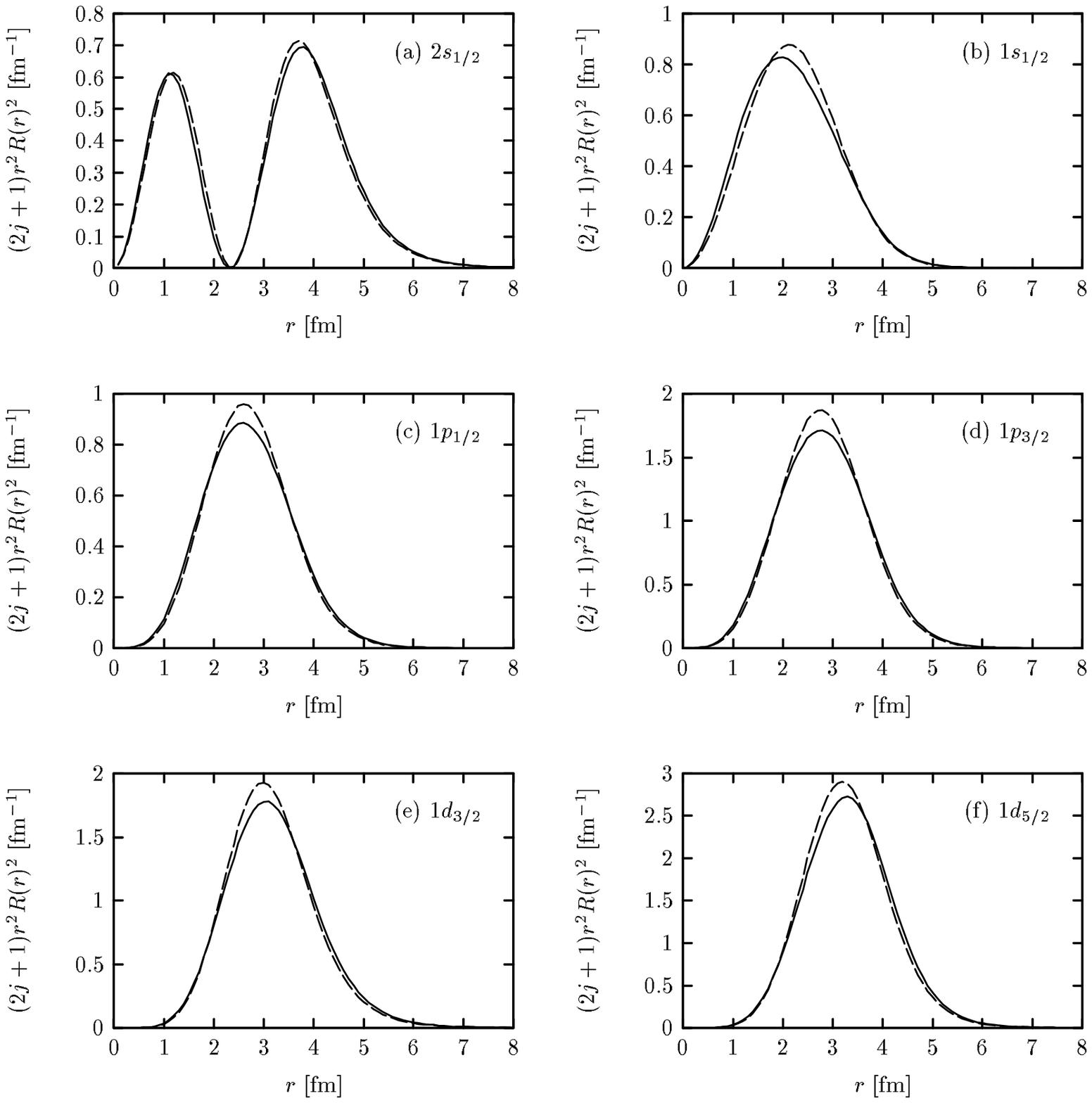}
\end{center}
\caption{\small
The same as fig. 7 for the nucleus $^{40}$Ca.
}
\end{figure}

The results for the correlated overlap functions of the nuclei
$^{12}$C, $^{16}$O and $^{40}$Ca are shown in figs. 7--9
respectively. In each one of these figures we show with solid lines
the radial density $r^2\psi(r)^2$ of the 
correlated overlap function and with dashed lines the
non-correlated result corresponding to the shell model.
We do not show the overlap functions of the 
also studied nucleus $^{48}$Ca.
Having stopped our calculation at 100 fm, it was not possible to obtain
convergence in this nucleus for the overlap function of the $p$-shell,
which requires higher values of $a$.

In figure 7 we show the first overlap functions of $^{12}$C for $l=0,1$.
In both cases  convergence is reached for relatively small values of 
$a=7$ and 8 fm, respectively, due to the fact that in the shell model
there are no second overlap functions  contributing to the OBDM. 
Short range correlations introduce extra contributions in the OBDM. 
However, these extra contributions decay  much faster 
than the exponential one, and very large distances are not needed to extract
the first overlap function. 
As we can see in figure 7, the inclusion of short-range
correlations produces in both cases 
a reduction of the maximum of the overlap
function
in coordinate space, while there is an increase for high $r$, which 
is better seen in the case of the $p$-shell, since it lies at higher
distances (panel (b)). 
In this last case we observe in addition that the overlap 
function   undergoes a small shift to the right due to the 
correlations.

%
%
\begin{table}[t]
\begin{center}
\begin{tabular}{llrcrr}\hline\hline
Nucleus     &  $nlj$       & $a$ [fm] & $|E_{c}-E_{ws}|/E_{ws}$ [\%]&
$S_{\rm th}$ & $S_{\rm exp}$\\\hline
$^{12}$C   & $1s_{1/2}$ &     7 & $4\times10^{-5}$ & 0.985 & 0.59 \\
           & $1p_{3/2}$ &     8 & $< 10^{-6}    $  & 0.986 & 0.56 \\\hline
$^{16}$O   & $1s_{1/2}$ &     7 & $2\times10^{-5}$ & 0.985 &      \\
           & $1p_{3/2}$ &       & $7\times10^{-4}$ & 0.986 & 0.59 \\
           & $1p_{1/2}$ &    86 & $4\times10^{-5}$ & 0.986 & 0.57 \\\hline
$^{40}$Ca  & $1s_{1/2}$ &       & $6\times10^{-2}$ & 0.988 & 0.75 \\
           & $2s_{1/2}$ &    14 & $2\times10^{-3}$ & 0.992 & 0.64 \\
           & $1p_{3/2}$ &       & $3\times10^{-2}$ & 0.985 & 0.72 \\
           & $1p_{1/2}$ &   100 & $3\times10^{-3}$ & 0.985 & 0.72 \\
           & $1d_{5/2}$ &       & $2\times10^{-3}$ & 0.985 & 0.74 \\
           & $1d_{3/2}$ &    64 & $3\times10^{-5}$ & 0.985 & 0.74 \\\hline
$^{48}$Ca  & $1s_{1/2}$ &       & $6\times10^{-4}$ & 0.986 &      \\
           & $2s_{1/2}$ &    16 & $1\times10^{-1}$ & 0.991 &      \\    
           & $1p_{3/2}$ &       & $2\times10^{-1}$ & 0.946 &      \\
           & $1p_{1/2}$ & $>100$& $3\times10^{-1}$ & 1.058 &      \\
           & $1d_{5/2}$ &       & $1\times10^{-3}$ & 0.983 &      \\
           & $1d_{3/2}$ &  100  & $6\times10^{-4}$ & 0.983 &      
\\\hline\hline
\end{tabular}
\end{center}
\caption{\small For each one of the quasi-hole states
in the closed shell nuclei studied we show: the asymptotic distance $a$
for which convergence of the correlated overlap function is reached
(third column), the relative difference between correlated and
uncorrelated separation energies (fourth column),
 and the computed spectroscopic factor (fifth column).
For comparison we show also the experimental value of the
spectroscopic factors extracted from $(e,e'p)$ experiments.  }
\end{table}

Similar effects are observed in figs. 8 for $^{16}$O and fig. 9 for 
$^{40}$Ca.
In all cases there is a reduction of the overlap  function at
intermediate distances (in most of them coinciding with the maximum of the
radial density)
and an increase for more large distances. 
As in $^{12}$C, we also observe a shift to the right of the 
overlap functions corresponding to the outer shells. This is the case
of the shells  $1p_{1/2}$ and $1p_{3/2}$ in $^{16}$O and 
$2s_{1/2}$, $1d_{3/2}$ and $1d_{5/2}$ in $^{40}$Ca. 

This shift effect over the outer shells can be understood in terms of
the repulsive nature of the NN interaction for short distances,
implicit in the correlation function $f(r)$, 
and the well known healing property of the wave function
for the two-nucleon system.
The correlation function produces a wound in
the $NN$ wave function $\Psi$, and what we are seeing in the overlap
function is the average effect of healing due to the interaction 
of the outer nucleons with the nucleons in the core.

However the inner shells do not show this effect because the
short-range repulsion due to the core partially cancels the one
produced by the external shells.  The net effect depends on the
particular nucleus and on the shell involved.  For instance, in the
case of the $1s_{1/2}$ shell in $^{40}$Ca, shown in panel (b) of
fig. 9, the correlations produce a shift of the overlap function to
the left, i.e., into the nucleus, since the short-range repulsion by
the outer shells tends to compress the $1s$ wave. The same compression
effect is observed in the inner lobe of the $2s_{1/2}$ (see panel (a)
of fig. 9).

In the cases of the intermediate shells $1p_{1/2}$ and $1p_{3/2}$ 
for $^{40}$Ca the joint effect of repulsion by the inner and outer
shells produces a shift to the left in the low $r$ region and a shift to the 
right for large $r$. Hence the net effect of correlations over these shells
is a widening of the overlap function, as seen in panels (c) and (d) of 
fig.~9.  

Once the overlap functions have been extracted, we can compute the
corresponding separation energy by a fit to a function $C
e^{-kr-\eta\ln kr}/r$ for large distances.  We have performed this fit
in the interval between 11 and 28 fm for the correlated and
uncorrelated overlap functions obtaining essentially the same
energies. The inclusion of the logarithm Coulomb phase is important in
this fit for protons, since it can modify the extracted energies in
more than 2 MeV in the case of $^{40}$Ca.  The relative difference
between correlated and uncorrelated energies is shown in the fourth
column of table 4.  In all cases the differences are less than 0.5 \%
even in the case of the $^{48}$Ca shells where convergence was still
not found for the overlap function.  These numerical results confirm
that short-range correlations do not change the mean field values of
the separation energies for quasi-hole states.

Results for the spectroscopic factors resulting from our model
are shown in the fifth column of table 4.  These have been computed
as the norm of the corresponding correlated overlap function
\begin{equation}
S = \langle\phi | \phi\rangle.
\end{equation}
As seen in table 4, all of the computed spectroscopic factors are
slightly less than one, being in most of the cases around $S\sim
0.985$.  The only exception found in table 4 is the value
$S_{p_{1/2}}=1.058$ for $^{48}$Ca, which is not a definitive number
since the asymptotic point $a=100$ fm used 
is not large enough to reach convergence in this case.

Our results indicate that short-range correlations of Jastrow-type
reduce the shell model occupation probability no more than 2\%. This
reduction is not enough to explain the experimental values extracted
from $(e,e'p)$ analysis shown in column 5 of table 4. This is in
agreement with other studies which report values similar to ours for
the spectroscopic factors.  Van Neck {\em et al.} found in
\cite{Van97} that central correlations generate a reduction around
1--2\% of the occupancy probability in $^{16}$O. More recently
Fabrocini and Co' \cite{Fab00} have computed overlap functions within
the FHNC/SOC theory and report values around 0.97--0.99 for the
spectroscopic factors with Jastrow correlations. Spin-isospin and
tensor correlations (not included in our calculation) produce
additional reduction of these values to $S\sim 0.86$--0.9 for the
valence shells.  The discrepancy with experimental values could be
further reduced by the inclusion of long-range correlations
\cite{Geu96}.  Center of mass corrections however produce an
enhancement of $\sim 7\%$ of the $p$-shell in $^{16}$O \cite{Van98a}.
Further investigation including all of these effects in a consistent
way is needed in order to clarify the situation.

\subsection{Quasi-particles and the continuum}

Up to now we have restricted our study to overlap functions
corresponding to quasi-hole states. Our correlated OBDM model 
allows us to compute also the multipoles $\rho_l(r,r')$ for high
values of $l$, which are expected to contain contributions coming from
quasi-particle states, i.e., states non occupied in the shell model
but which are partially populated in the ground state due to nuclear
correlations.  We have investigated if the overlap functions for
quasi-particles can be extracted from our OBDM using the asymptotic
method.

This study is motivated by a recent calculation done in \cite{Sto96}
where results for the quasi-particle overlap function for the
$1d$-shell in $^{16}$O and for the $1f$ shell in $^{40}$Ca are
presented. These authors begin with an OBDM including Jastrow
correlations and apply an asymptotic procedure similar to fit II in
order to extract the overlap functions.  For instance they report a
value of $S=0.01$ for the spectroscopic factor of the $1f$ shell in
$^{40}$Ca.  Neither the asymptotic interval used for the fit nor the
convergence distance are indicated in \cite{Sto96}. Apparently they
should not have used very high asymptotic values since they use a
harmonic oscillator basis that fall off rapidly at large distances.

However these results were criticized in ref. \cite{Van97}, where it
was shown that it is not possible, starting from a CBF-type wave
function, to generate bound-state overlap functions with quantum
numbers that are unoccupied in the Slater determinant.  The reason is
that the overlap functions decay exponentially with the same decay
constant as the hole single-particle orbital

In fact, when we apply the $\sqrt{\rho}$ method to compute a
quasi-particle overlap function from our radial OBDM for high values
of $l$ we do not obtain convergence within the 100 fm range and instead
the results decrease rapidly, being negligible for high $r$. We must
conclude that in our model it is not possible to obtain such
quasi-particle states.

This result can be understood  by examining the following
asymptotic expression of the OBDM for unoccupied multipole $l$
\begin{equation}
\rho_{l}(r_1,r'_1)
\sim
C f(r_1,r'_1),
\kern 1cm
r_1,r'_1 \rightarrow\infty
\end{equation}
where $C$ is a constant and the function $f(r,r')$ is given by
\begin{equation}\label{aproximacion-asintotica}
f(r_1,r'_1) = 
\frac{ \exp{(-2kr_1)}\exp{(-2kr'_1)}
       \exp\left( \frac{-B (r_1-r'_1)^2 }{2} \right)
     }
     { r_1^2 r'_1{}^2 (r_1+r'_1)^2   
     }.
\end{equation}
Here $k$ is the wave number of the (occupied) valence shell
and $B$ is the parameter of the correlation function.
This expression is proved in Appendix B for the simplest case of the
multipole $l=2$ for the nucleus $^{12}$C.

In figure 10 we show with solid lines the computed radial density
$\rho_2(r,r')$ for $^{12}$C as a function of $r$ for several fixed
values of $r'$.  In addition we show with dashed lines the function
$f(r,r')$ multiplied by a convenient constant $C$ fitted to the
density.  We see that in fact the above asymptotic expression is
approximately verified by the computed density for high values of $r$
and $r'$.

\begin{figure}[t]
\begin{center}
\leavevmode
\epsfbox[100 525 500 730]{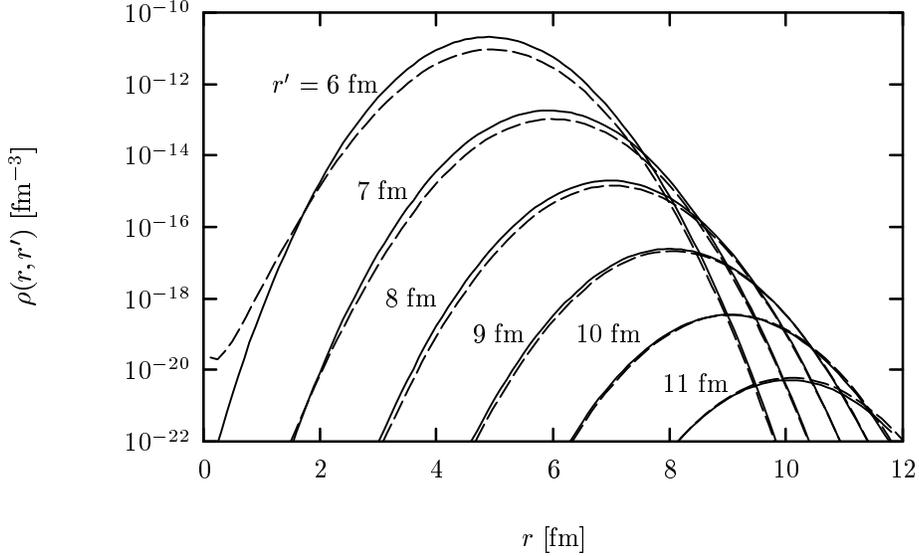}
\end{center}
\caption{\small The solid lines are the correlated OBDM,
$\rho_{2}(r,r')$ for $l=2$ in $^{12}$C, as a function of $r$ for
different values of $r'=6$, 7, 8, 9, 10 y 11 fm.  The dashed lines are
a fit to an asymptotic approximation $f(r,r')$ defined in
(\protect\ref{aproximacion-asintotica}).  }
\end{figure}

If we now try to compute a quasi-particle  overlap function using the 
$\sqrt{\rho}$ method we obtain 
\begin{equation}
\psi_{l}(r) 
=
\frac{\rho_{l}(r,a)}{\sqrt{\rho_{l}(a,a)}}
\nonumber\\
\sim
\frac{2a}{r^2(r+a)^2} \rho_{0}(r) e^{-B(r-a)^2/2}
\end{equation}
this expression goes to zero for $a \rightarrow \infty$, since the
diagonal part of $\rho_{l}$ has an exponential decay, while the
non-diagonal part has an additional Gaussian behavior corresponding to
the correlation function. This explains why in our results the
extracted overlap function for quasi-particles are zero.  Thus in our
model of correlated OBDM it is not possible to obtain the overlap
functions for quasi-particles since the corresponding information of
single-particle states above the Fermi level (configuration mixing)
has not been included into the model.  Moreover, when one subtract
from the correlated OBDM the contribution of the quasi-hole states the
remaining density contains only the contributions coming from the {\em
continuum} states of the residual nucleus.  These contributions are
implicit in the expansion (\ref{densidad-overlap}) in terms of overlap
functions and they can be expressed as an integral over the energy.
\begin{equation}\label{continuum}
\rho(\nr,\nr')-\rho_{quasi-hole}(\nr,\nr')= 
\int dE \Psi_E(\nr)^{\dagger}\Psi_E(\nr')
\end{equation}
With the asymptotic method studied here it is not possible to extract
these overlap functions of the continuum. For this a practical
inversion method of the integral (\ref{continuum}) in the asymptotic
region is needed.  The knowledge of these overlap functions would be
of interest, for instance, to compute the $(e,e'p)$ cross section for
high missing energy.

\subsection{Exclusive response functions and cross sections}

In figs. 11--13 we show the five exclusive responses for proton
knockout from the valence shells of the nuclei $^{12}$C, $^{16}$O and
$^{40}$Ca, as a function of the missing momentum.  In all the cases
the kinematics correspond to a fixed value of the momentum transfer
$q=460$ MeV/c and $\omega$ fixed around the quasi-elastic peak.  In
each panel of figs. 11--13 we show four curves corresponding to
different models for the initial and/or final wave functions that
enter in the current matrix element (\ref{hadronic-tensor}).  We show
results for PWIA and DWIA with and without short-range correlations in
the initial state overlap functions.  The DWIA results
have been obtained using for the FSI the optical potential of
ref. \cite{Sch82}.  Specifically, the solid lines have been computed
with the DWIA model and using correlated overlap functions, while the
dashed lines do not include correlations. Thus comparison between
solid and dashed lines shows the effect of short-range correlations in
the responses.  Results in PWIA with and without correlations are
shown with short-dashed and dotted lines, respectively.

\begin{figure}[ph]
\begin{center}
\leavevmode
\epsfbox[100 270 500 750]{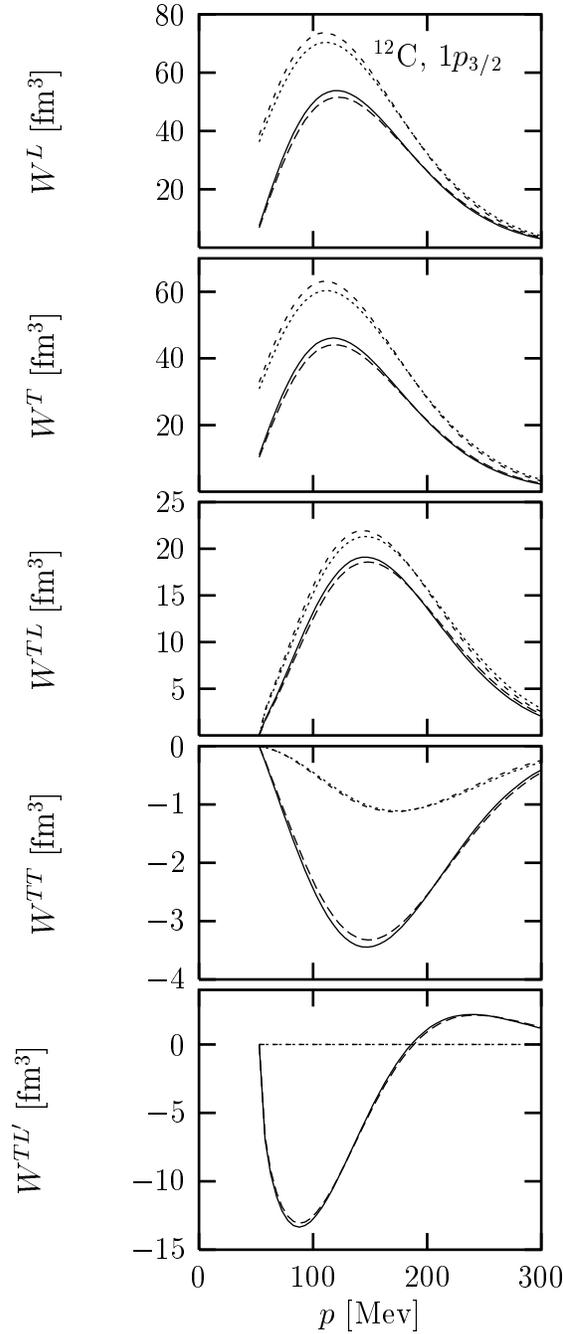}
\end{center}
\caption{\small Response functions for proton knock-out from the
$p_{3/2}$ shell in $^{12}$C, for $q=460$ MeV and $\omega$ at the
quasi-elastic peak. Results are shown in DWIA with (solid lines) and
without (dashed lines) correlations in the hole overlap function, and
in PWIA with (short-dashed lines) and without (dotted lines)
correlations.  }

\end{figure}

\begin{figure}[ph]
\begin{center}
\leavevmode
\epsfbox[100 230 500 710]{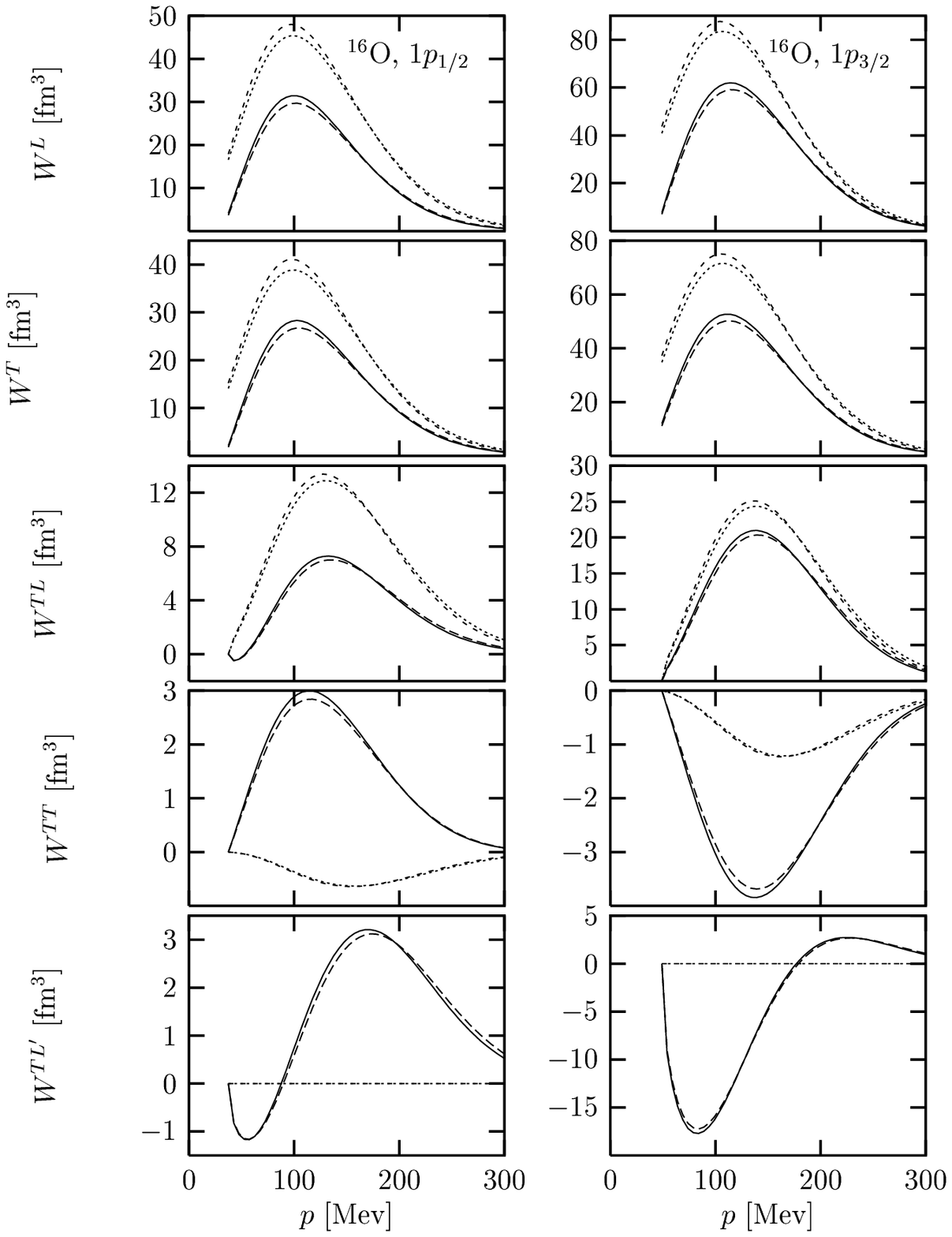}
\end{center}
\caption{\small Response functions for proton knock-out from the
valence shells, $1p_{3/2}$ and $1p_{1/2}$, in $^{16}$O, for $q=460$
MeV and $\omega$ at the quasi-elastic peak. The meaning of the lines
is the same as in fig. 11.  }
\end{figure}

In all the cases we note in the region $p < 200$ MeV/c an increase of
the L, T and TL responses due to correlations, which is around 5\%
near the maximum. This increase is quite independent on the FSI since it
is also present in PWIA. The reason for this fact is that correlations 
between the ejected proton and the residual nucleus 
 have not been included.  The increase seen in the
responses can be easily understood in PWIA, where we are basically seen the
momentum distribution of the shell,  as a consequence of the
hardening effect
seen in the overlap functions in figs. 7--9 for the valence shells.
 Since
in momentum space the low $p$ region is sensitive to the high $r$
region, the increase of the overlap function for high $r$ translates
into an increase of the Fourier transform for low $p$ ($\sim 100$
MeV/c) where the maximum of the momentum distribution is located.
Since the correlations in the ground state are in some degree 
decoupled from the FSI, the same effect is propagated to the case of
the  DWIA.

In the case of the TT response, we also find an increase of its
absolute value due to correlations in DWIA, which is not seen in PWIA
because the leading-order magnetization current do not contribute to
this response \cite{Ama98a}, and the resulting factorized
single-nucleon TT response is of order $(p/M)^2$ in a non-relativistic
expansion (this is the reason why this response function is so small).
This kinematical dependence comes exclusively from the convection
current, producing a hardening of the maximum of the momentum
distribution toward higher $p$-values $\sim 150$ MeV, where the
correlated and uncorrelated results are closer.  In DWIA, the FSI
breaks the factorization property and the magnetization current gives
a contribution which wherefore is much larger than the PWIA result.

Regarding the fifth response function TL$'$, which only can be
measured using polarized electrons, it is exactly zero in absence of
FSI.
In DWIA however it produces a contribution to the total cross section and
the correlations in the ground state produce an increase which is in
general of the same
order  as in the unpolarized responses. This increase is
 even more large ($\sim 15\%$)  in the case of the 
$1d_{3/2}$ shell in $^{40}$Ca (see fig. 13).

\begin{figure}[ph]
\begin{center}
\leavevmode
\epsfbox[100 230 500 710]{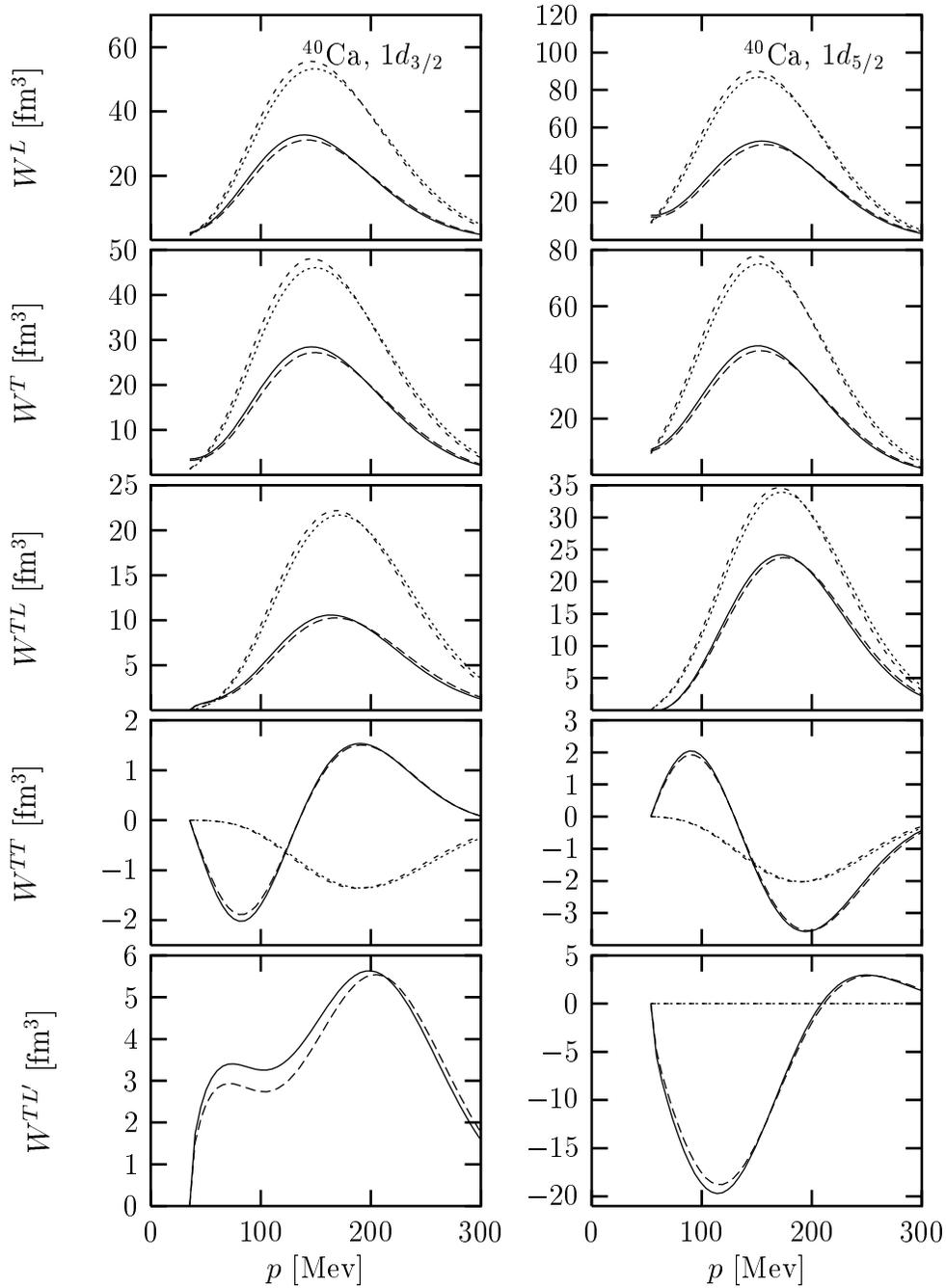}
\end{center}
\caption{\small
Response functions for proton knock-out from the valence shells,
$1d_{5/2}$
and $1d_{3/2}$, in $^{40}$Ca,
for $q=460$ MeV and $\omega$ at the quasi-elastic peak. The meaning of
the lines is the same as in fig. 11.
}
\end{figure}

State-independent short-range correlations produce an increase of the
$(e,e'p)$ cross section, since the later is a linear combination of
the several response functions appearing in eq. (1).  An example is
shown in fig. 14, where results of DWIA calculations for the reaction
$^{16}$O$(e,e'p)$ are displayed together with the experimental data of
ref. \cite{Chi91}.  Here the kinematics correspond to fixed momentum
transfer $q=570$ MeV/c and energy transfer $\omega=170$ MeV, at the
quasi-elastic peak. The energy of the electron beam is 580 MeV and the
proton is emitted in the scattering plane.  In fig. 14 we show with
solid lines our DWIA results using the correlated overlap functions
for the $1p_{1/2}$ and $1p_{3/2}$ shells, while with dashed lines we
show the uncorrelated results.  We note an enhancement of the cross
section due to correlations which is of the same order of magnitude as
was found for the response functions, and which clearly increases the
disagreement between theory and experiment.  In the same figure we
show with short dashed and dotted lines the computed cross sections
multiplied by the factors 0.6 ($1p_{1/2}$) and 0.5 ($1p_{3/2}$) in the
correlated case and 0.64 ($1p_{1/2}$) and 0.53 ($1p_{3/2}$) in the
uncorrelated one.  

\begin{figure}[t]
\begin{center}
\leavevmode
\epsfbox[100 370 500 710]{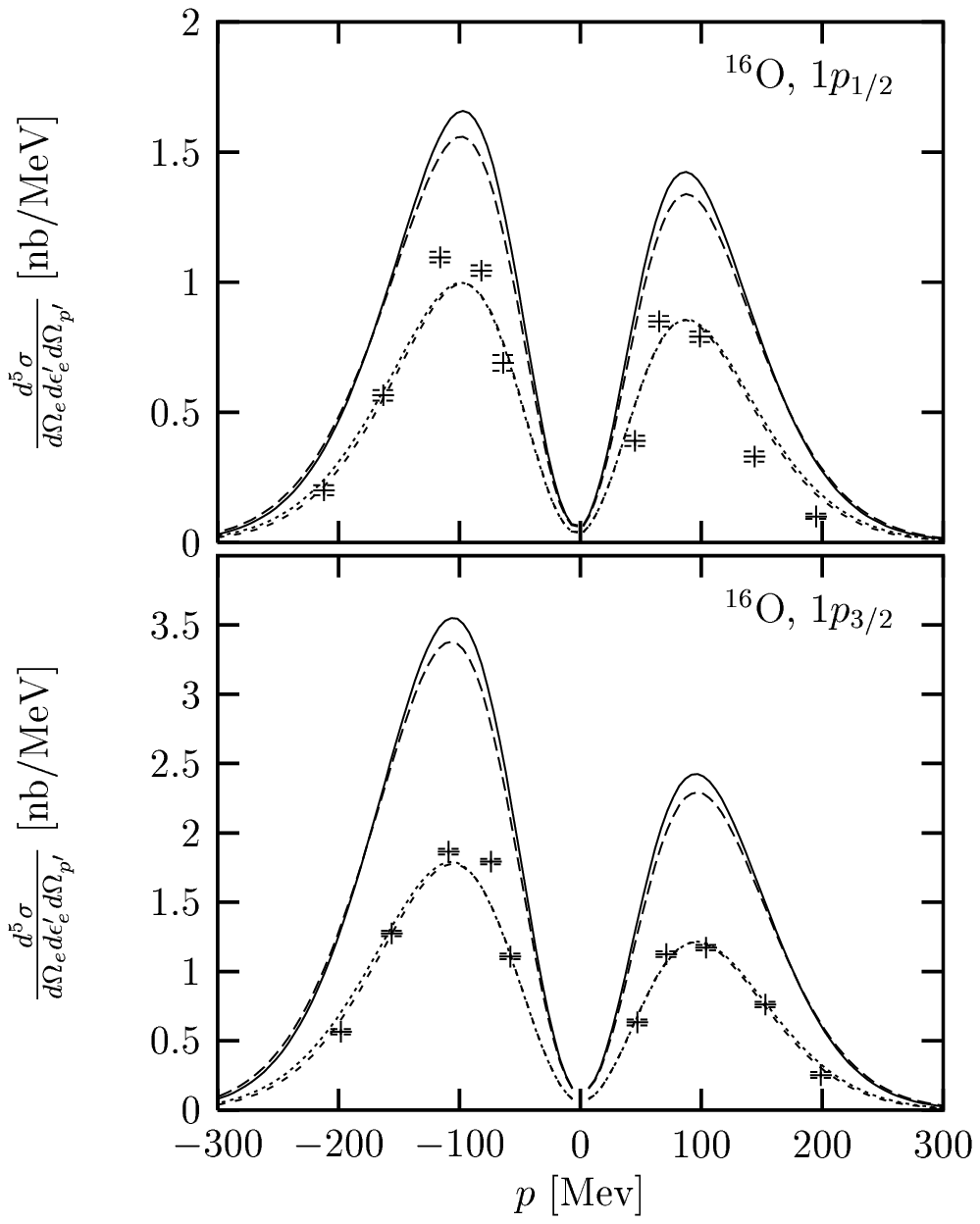}
\end{center}
\caption{\small
Computed $(e,e'p)$  cross section for the valence shells
of $^{16}$O. The solid lines include correlated overlap functions 
while the dashed lines do not. These calculations have been scaled to
give the dashed and dotted lines in order to reproduce the
experimental data from ref. \protect\cite{Chi91}.}
\end{figure}

Hence the scaling factor needed to reproduce the experimental cross
section is {\em smaller} for correlated than for uncorrelated overlap
functions even though the computed spectroscopic factors for these
shells are smaller than one, $S= 0.985$ (see table 4).  This fact does
not necessarily imply a decrease of the {\em experimental}
spectroscopic factors since these are obtained by a
simultaneous  fit of the parameters of the single particle potential
also. In other words, experimentally one searches for 
the best phenomenological
overlap function that when included into an uncorrelated DWIA code 
reproduces the experimental data.
Our model has not adjustable parameters since the correlations are
already included and so are the spectroscopic factors. 
Our results are showing that short-range correlations of the
central-type in the ground state do not produce an improvement of the
$(e,e'p)$ data description.  
Tensor correlations and long-range correlations are obvious candidates
for a reconciliation between theory and experiment.

\begin{figure}[t]
\begin{center}
\leavevmode
\epsfbox[100 370 500 710]{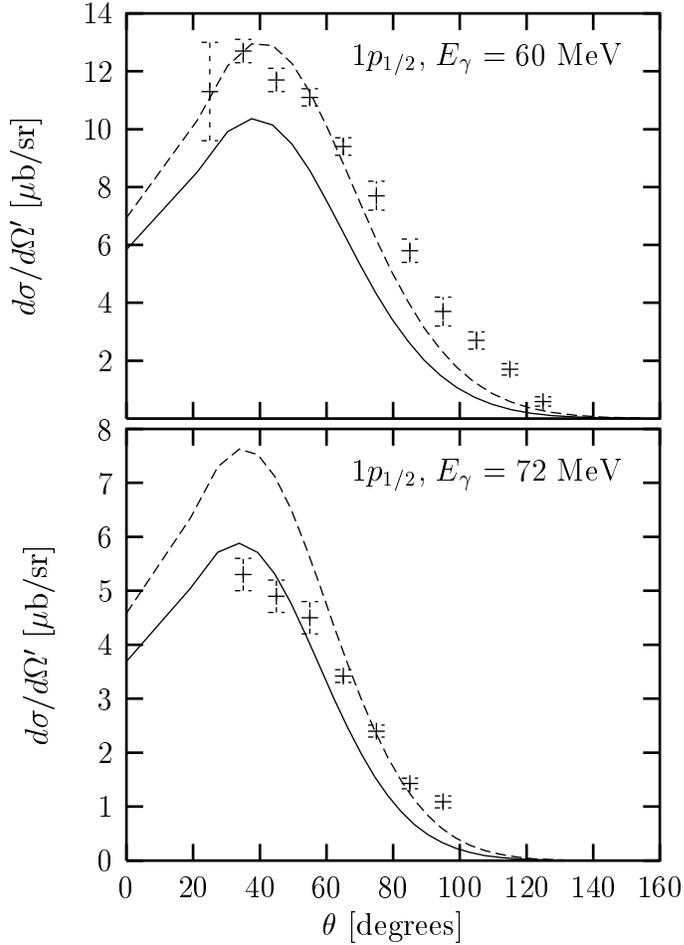}
\end{center}
\caption{\small
Computed $(\gamma,p)$ cross section for the $1p_{1/2}$ shell
of $^{16}$O. The solid lines include correlated overlap functions
while the dashed lines do not. Experimental data are from ref. 
\protect\cite{Mil95}} 
\end{figure}

To end the discussion we give in fig. 15 another application of our
 DWIA model in photo-nuclear reactions.  Therein we show the computed
 $(\gamma,p)$ cross section  from the  $1p_{1/2}$ shell in $^{16}$O for
 two beam energies of $E_{\gamma}=60$ and 72 MeV together with the
 experimental data of ref. \cite{Mil95}. 
Again we show with solid and dashed
 lines the correlated and uncorrelated results respectively. No
 scaling factors are included. The
 impact of central
 correlations in this case is completely different from the $(e,e'p)$
 reaction. In this case they produce a large reduction of the cross section. 
The difference between the two reactions lies in the different
 kinematical region which is being probed by photons. 
In the case of $(\gamma,p)$ the energy-momentum
transfer verifies $\omega=q$ and we are far from the quasi-elastic peak
 region.
As a consequence the missing momentum is well above 200 MeV/c.
Thus apart from the small values of $q$, the photon is exploring here
 the high momentum tail of the overlap function where correlations 
produce a reduction of the momentum distribution. 
Such reduction can also be appreciated in figs. 11--13, where the 
correlated transverse response (the one contributing to photo-reactions) 
is below the uncorrelated one for $p>200$ MeV/c.
One should also be aware of the difficulties presents in the DWIA
description of the $(\gamma,p)$ reaction for so low energies and 
high missing momenta, where  in particular 
the orthogonality approximation (10) 
is no longer true and other effects \cite{Joh00}  of the same order 
as the ones arising from correlations 
could appreciably change the results of fig. 15.

\section{Summary and Conclusions}

In this work the effects of short-range correlations on $(e,e'p)$
observables and overlap functions have been studied.  The starting
point for the present calculation has been the OBDM, including
short-range correlations of central Jastrow type, and which has been
computed by a cluster expansion to leading order in the correlation
function.  Correlated overlap functions corresponding to quasi-hole
states in closed-shell nuclei have been extracted from the asymptotic
OBDM multipoles, $\rho_l(r,r')$, computed at large asymptotic
distances $r'\leq 100$ fm.  The reliability of the extraction method
has been tested in the shell model, where a detailed study of the
different fit procedures and of the convergence distances has been
performed. These distances have been found to be very large in the
cases in which two single-particle states of close energies are
present.  In those cases the asymptotic distances considered in our
calculation of the OBDM have been enough to separate the corresponding
overlap functions of the nuclei $^{12}$C, $^{16}$O and $^{40}$Ca.  As
also has been found in other works \cite{Ari97,Fab00,Fab00b},
short-range correlations produce small effects on the density
distribution and likewise on the OBDM. Our results given in
figs. 7--9 show that these effects are more noticeable in the overlap
functions since they are determined by the asymptotic behavior of the
OBDM.  Short-range correlations produce a redistribution of the
single-particle densities in coordinate space. Their values are
reduced at the maximum and increased for large distances.  In the case
of the valence shells we find a hardening of the distribution 
related to the collective effect of the NN repulsion at short distances.

The values of the computed spectroscopic factors in the present work
are around 0.985, in accord with the findings obtained with other
techniques. It is known that tensor and long-range correlations can
reduce these values but, at present, no model is able to
reproduce the experimental values extracted from $(e,e'p)$ data.

The computed overlap functions have been included in a DWIA model of
the $(e,e'p)$ reaction, and exclusive response functions and cross
sections have been computed for quasi-elastic kinematics. Although the
computed spectroscopic factors are less that one, we have found an
increase ($\sim 5\%$) of the response functions and, accordingly, of
the cross section in the region of the maximum of the missing momentum
distribution for knockout from the valence shells.  This reduction is
independent on the FSI and is a consequence of the increase of the
single-particle densities for large distances.  Thus the inclusion of
central correlations is worsening the description of the experimental
data in our model.  This again proves that central correlations alone
are not enough to describe this reaction successfully. Apart from
spin-isospin and tensor correlations, not included here for
computational reasons, we would like to remark the necessity of a
model including in addition long-range correlations in a consistent
way.

The later correlations are related to the presence of admixture of
multi-$\hbar\omega$ configurations into the valence wave functions of
the residual nucleus \cite{Blu85}.  For instance in the case of the
$^{40}$Ca nucleus, the residual states correspond to the nucleus
$^{39}$K. It is known that the transverse form factors of the
measured elastic and inelastic transitions in $^{39}$K show
significant departures from the single-particle picture and that a
modification of the extreme shell model wave functions through the
effect of core polarization is needed to describe the electron
scattering cross section \cite{Blu85}.  When computing the $(e,e'p)$
reaction in $^{40}$Ca one uses a wave function which reproduces the
elastic electron scattering data or equivalently, the charge density,
corresponding to the initial state (the ground state of $^{40}$Ca),
ignoring the necessity of a proper description of the residual states
also, usually treated as single holes in the core.

Concerning the $(\gamma,p)$ reaction, central correlations play here a
more important role, producing a large reduction of the cross section,
since this reaction is sensitive to the high momentum components of
the wave function where correlation effects are maximized.

In this paper we have demonstrated with a realistic model that the
asymptotic method to compute the OBDM is feasible and that convergence
of the results can be obtained: therefore it rises as an alternative,
reliable starting point to be applied to other kind of correlated
densities.

\subsection*{Acknowledgments}

We would like to thank G.\ Co' for  fruitful discussions.
This work was  supported by funds provided by 
DGICYT (Spain) under contracts PB/98-1367, PB/98-1318,
and by the Junta de Andaluc\'{\i}a (FQM225, FQM220).

\appendix

\section{Multipole analysis of exclusive response functions}

The general multipole analysis of the $(e,e'p)$ responses including
polarization degrees of freedom of the target nucleus and electron was
presented in refs.~\cite{Ama98a,Ras89}.  The formalism can also be
applied to the present case of  $J=0$ nuclei, where some
simplification of the multipoles given in ref. \cite{Ama98a} can be done. 
The following equations  we write in this appendix 
have been obtained, after some work, from the
corresponding equations of ref. \cite{Ama98a} for the particular
case ${\cal J}=0$ (${\cal J}$ being in \cite{Ama98a} 
the angular momentum corresponding
to a multipole expansion in terms of spherical harmonics 
$Y_{\cal JM}(\theta^*,\phi^*)$ of the nuclear polarization angles). 

We expand the nuclear electromagnetic current as a sum  of 
Coulomb (for the time component), and electric and
magnetic  (for the transverse three-vector current) 
multipole operators of rank $J$.
The final hadronic state is  also expanded in
partial waves of the ejected proton, as a combination of hadronic
states with total angular momentum $J$, 
 denoted as $|\sigma\rangle \equiv |(lj)J_{\alpha};J\rangle$,
that represents a nucleon in the continuum with asymptotic
angular momenta $lj$ coupled with the  residual nuclear state
$|\Phi^{(A-1)}_{\alpha}\rangle=|J_{\alpha}\rangle$. The exclusive
response functions can be written in the form
\begin{eqnarray}
W^L     &=& \frac{1}{K}
             \sum_{L\geq 0} [L]P^0_L
             (\cos\theta')W_L^{L} \label{eq2}\\
W^T     &=& \frac{1}{K}
              \sum_{L\geq 0} [L]P^0_L
              (\cos\theta')W_L^{T} \label{eq3}\\
W^{TL}  &=& -\frac{1}{K}
              \sum_{L\geq 1} \frac{[L]}
              {\sqrt{L(L+1)}}
               P^1_L(\cos\theta')W_L^{TL} \label{eq4}\\
W^{TL'} &=&  \frac{1}{K} \sum_{L\geq 1}
              \frac{[L]}{\sqrt{L(L+1)}}
              P^1_L(\cos\theta')W_L^{TL'} \\
W^{TT}  &=& \frac{1}{K}
            \sum_{L\geq 2} \frac{[L]}
            {\sqrt{(L-1)L(L+1)(L+2)}}
             P^2_L(\cos\theta')W_L^{TT}, 
                        \label{eq5}
\end{eqnarray}
with $[L]=\sqrt{2L+1}$. Note that the whole dependence on
the emitted proton angle $\theta'$ is given through the Legendre
functions $P_L^M(\cos\theta')$ and that the present response functions  
are  divided by the factor  $K=Mp'/(2\pi\hbar)^3$ respect to those of
ref. \cite{Ama98a}. The reduced response functions $W^K_L$, defined as
the coefficients in the expansions (\ref{eq2}--\ref{eq5}) are given by
\begin{eqnarray}
W^{L}_L &=& \sum_{\sigma'\sigma}
                        \Phi_{\sigma'\sigma}(L)
                        \tresj{J}{J'}{L}{0}{0}{0}
                        \xi^+_{J'-l',J-l} 
                        R^L_{\sigma'\sigma}
                        \label{eq6}\\
W^{T}_{L} &=& -\sum_{\sigma'\sigma}
                        \Phi_{\sigma'\sigma}(L)
                        \tresj{J'}{J}{L}{1}{-1}{0}
                        (\xi^+_{J'-l',J-l} 
                        R^{T1}_{\sigma'\sigma}
                        +\xi^-_{J'-l',J-l} 
                        R^{T2}_{\sigma'\sigma})
                        \\
W^{TL}_{L} &=& -2
                        \sum_{\sigma'\sigma}
                        \Phi_{\sigma'\sigma}(L)
                        \tresj{J'}{J}{L}{0}{1}{-1}
                        (\xi^+_{J'-l',J-l} 
                        R^{TL1}_{\sigma'\sigma}
                        -\xi^-_{J'-l',J-l} 
                        R^{TL2}_{\sigma'\sigma})
                        \\
W^{TL'}_{ L} &=& -2
                        \sum_{\sigma'\sigma}
                        \Phi_{\sigma'\sigma}(L)
                        \tresj{J'}{J}{L}{0}{1}{-1}
                        (\xi^+_{J'-l',J-l} 
                        I^{TL1}_{\sigma'\sigma}
                        -\xi^-_{J'-l',J-l} 
                        I^{TL2}_{\sigma'\sigma})
                        \\
W^{TT}_{L} &=& -\sum_{\sigma'\sigma}
                        \Phi_{\sigma'\sigma}(L)
                        \tresj{J'}{J}{L}{1}{1}{-2}
                        (\xi^+_{J'-l',J-l} 
                        R^{TT1}_{\sigma'\sigma}
                        -\xi^-_{J'-l',J-l} 
                        R^{TT2}_{\sigma'\sigma})
                        \label{eq9}.
\end{eqnarray}
Note that the $TL$ and $TL'$ reduced response functions 
of ref. \cite{Ama98a} include an extra factor $\sqrt{2}$ due to the
different definition of the $v_{TL}$ and $v_{TL'}$ factors.
The coupling coefficient $\Phi_{\sigma'\sigma}(L)$ includes the
internal sums over third components and it is defined as
\begin{equation}
\Phi_{\sigma'\sigma}(L)= P^+_{l+l'+L}
        [J][J'][j][j'][L]
        (-1)^{J+J'+J_{\alpha}+1/2+L}
        \tresj{j'}{j}{L}{\textstyle\frac12}{-\textstyle\frac12}{0}
        \seisj{j'}{j}{L}{J}{J'}{J_{\alpha}}
        \label{e-26}
\end{equation}
We also use the parity functions
\begin{eqnarray}
P^{\pm}_i   &=& \frac12[1\pm(-1)^i] \\
\xi^+_{ij}  &=& (-1)^{(i-j)/2}P^+_{i+j} \\
\xi^-_{ij}   &=& (-1)^{(i-j+1)/2}P^-_{i+j}.
\end{eqnarray}
In order to define the functions $R^K_{\sigma',\sigma}$ and
$I^K_{\sigma',\sigma}$ in eqs.~(\ref{eq6})--(\ref{eq9}), 
we introduce the
Coulomb, electric and magnetic multipole matrix elements
\begin{eqnarray}
C_{\sigma} &\equiv& \langle \sigma \| \hat{M}_J(q) \| 0\rangle
\label{eq14}\\
E_{\sigma} &\equiv& \langle \sigma \| \hat{T}^{el}_J (q) \| 0\rangle
\\
M_{\sigma} &\equiv& \langle \sigma \| i\hat{T}^{mag}_J (q) \| 0\rangle.
\label{eq16}
\end{eqnarray}
where $\hat{M}_J(q)$, $\hat{T}_J^{el}(q$) and $\hat{T}_J^{mag}(q)$ are
the usual Coulomb, electric and magnetic multipole operators.  The
functions $R^K_{\sigma',\sigma}$ and $I^K_{\sigma',\sigma}$ in
eqs.~(\ref{eq6})--(\ref{eq9}) are then defined by the following
quadratic forms constructed with these multipoles
\begin{eqnarray}
R^L_{\sigma'\sigma}&=&
{\rm Re}\left[C^*_{\sigma'}C_{\sigma}\right] 
\label{eq17}\\
R^{T1}_{\sigma'\sigma}&=&
{\rm Re}\left[E^*_{\sigma'}E_{\sigma} +
        M^*_{\sigma'}M_{\sigma}\right]
\label{eq18} \\
R^{T2}_{\sigma'\sigma}&=&
{\rm Re}\left[E^*_{\sigma'}M_{\sigma} -
        M^*_{\sigma'}E_{\sigma}\right]
\label{eq19} \\
R^{TL1}_{\sigma'\sigma}&=&
{\rm Re}\left[C^*_{\sigma'}E_{\sigma}\right]
\label{eq20} \\
R^{TL2}_{\sigma'\sigma}&=&
{\rm Re}\left[C^*_{\sigma'}M_{\sigma}\right]
\label{eq21} \\
I^{TL1}_{\sigma'\sigma}&=&
{\rm Im}\left[C^*_{\sigma'}E_{\sigma}\right]
\label{eq20b} \\
I^{TL2}_{\sigma'\sigma}&=&
{\rm Im}\left[C^*_{\sigma'}M_{\sigma}\right]
\label{eq21b} \\
R^{TT1}_{\sigma'\sigma}&=&
{\rm Re}\left[E^*_{\sigma'}E_{\sigma} -
        M^*_{\sigma'}M_{\sigma}\right]
\label{eq22} \\
R^{TT2}_{\sigma'\sigma}&=&
{\rm Re}\left[E^*_{\sigma'}M_{\sigma} +
        M^*_{\sigma'}E_{\sigma}\right].
\label{eq23} 
\end{eqnarray}
The $L$, $T$, $TL$ and $TT$  responses  include only the real
parts of the quadratic combinations of the various multipole matrix
elements, while the fifth response function $TL'$ 
is a linear combination of the imaginary parts (\ref{eq20b},\ref{eq21b}).
Therefore the $TL'$ response
is zero in PWIA, where the matrix  elements (\ref{eq14}--\ref{eq16})
are real numbers. In presence of an interaction, however, the matrix 
elements (\ref{eq14}--\ref{eq16})
are in general complex  numbers, due to the asymptotic complex 
phase ${\em e}^{i\delta_{lj}}$ introduced by the nuclear interaction
in the wave function, and as a consequence, the fifth response
function is different from zero in DWIA.  
The sum over the quantum numbers
$\sigma=(l,j,J)$, $\sigma'=(l',j',J')$ and $L$
in eqs.~(\ref{eq2})--(\ref{eq9}) is
only restricted by angular momentum conservation.
In practical calculations we fix the number of multipoles involved in the
sums by comparing our results with the ones corresponding to the
factorized PWIA in the impulse approximation
where, as known, the nuclear responses can be
computed exactly. 

The outgoing proton wave function corresponds to a solution
of the Schr\"odinger equation for
positive energies using a complex optical potential fitted
to elastic proton-nucleus scattering data.
The partial wave $l j$ with energy $E>0$ and
wave number $k=\sqrt{2ME}$ is determined by the
asymptotic condition
\begin{equation}
R_{l j}(k,r)\sim \sqrt{\frac{2M}{\pi\hbar^2k}}
    {\rm e}^{-i(\sigma_l+\delta^*_{l j})}
    \sin\left( kr-\eta\log2kr-l\frac{\pi}{2}+
               \sigma_l+\delta_{l j}^*
         \right)
\end{equation}
where $\delta_{l j}$ is the complex phase-shift 
and $\sigma_l$ is the
Coulomb phase-shift. In the limit in which the imaginary part of 
the  optical potential is zero, the phase-shift
$\delta_{l j}$ a real number, the continuum radial
wave functions are normalized with a Dirac delta function containing the
energies (see Ref.~\cite{Ama98a}).
The imaginary (absorptive) part of the optical potential modifies the
normalization of the continuum sates since the flux of the
outgoing particles in the elastic channel is reduced.


\section{Asymptotic OBDM for unoccupied states}


We consider as an example the simplest case of the multipole $l=2$ for
the nucleus $^{12}$C. More details are given in \cite{Maz01}.
 We use the expression
(\ref{densidad-correlacionada}) for the correlated contributions to
the OBDM.  First we exclude the contribution of diagrams C and D of
fig.~2, since the dependence of these terms on the density coordinates
$\nr,\nr'$ is done across the non correlated density
$\rho_0(\nr,\nr')$ which only contains the multipoles $l=0,1$ in the
case of $^{12}$C. In other words, the external points $\nr$ and $\nr'$
in diagrams C and D are connected with the others with density lines
only, which cannot modify its multi-polarity $l=0,1$.

In  the case of diagram A of fig. 2, it  can be written in a multipole
expansion as \cite{Ari97}
\begin{eqnarray}
\rho_A(\nr_1,\nr'_1)
&=&
\sum_{n_1 l_1 j_1}
(2j_1+1)R_{n_1 l_1 j_1}(r_1) R_{n_1 l_1 j_1}(r'_1) 
\sum_{l l_2 }
\frac{2l_3+1}{2l_2+1}
\left( \begin{array}{ccc}
            l_1 & l_2 & l \\
             0  &  0  & 0 
       \end{array}
\right)^2
P_l(\cos\theta_{11'})
\nonumber\\
&& 
\mbox{}\times
\int_0^\infty dr_2\, r_2^2
\rho_0(r_2)
f_{l_2}(r_1,r_2)
f_{l_2}(r'_1,r_2)
\end{eqnarray}
where the function $f_{l_2}(r_1,r_2)$ is the multipole of the
correlation function for angular momentum $l_2$.  The sum over $n_1,
l_1, j_1$ corresponds to the occupied states $1s_{1/2}$ e $1p_{3/2}$
in $^{12}$C.  The multipole $l=2$ of the OBDM is obtained as the
coefficient of the Legendre polynomial $P_2(\cos\theta_{11'})$ in the
above equation.  Since we are interested in the asymptotic behavior
for $r_1,r'_1\rightarrow\infty$ we only consider the contribution
coming from the $l_1=1$ term, i.e., $(n_1 l_1 j_1)= 1p_{3/2}$.  Hence
the 3-$j$ coefficient gives a non zero result for $l_2=1$ only.  The
corresponding multipole $l_2=1$ for the correlation function is
proportional to the integral
\begin{equation}
\int d\cos\theta_{12}\, \cos\theta_{12} f(\nr_{12})
=
-A \int d\cos\theta_{12} \cos\theta_{12} e^{-Br_{12}^2}
\sim
-\frac{A}{2Br_1 r_2} e^{-B(r_1-r_2)^2}.
\end{equation} 
Hence the asymptotic behavior of the $l=2$ multipole for
$r_1,r'_1\rightarrow\infty$ is 
\begin{equation}
\rho_{A}(r_1,r'_1)_{l=2}
\sim
R_{1p_{3/2}}(r_1)
R_{1p_{3/2}}(r'_1)
\int_0^\infty dr_2\, r_2^2
\rho_0(r_2)
\frac{ e^{-B(r_1-r_2)^2}}{r_1 r_2}
\frac{ e^{-B(r'_1-r_2)^2}}{r'_1 r_2}.
\end{equation}
Changing to the variable  $r'_2=r_2-r_m$, where
$r_m=(r_1+r'_1)/2$
is the mid point between 
$r_1$ and  $r'_1$, we arrive to 
\begin{equation}
\rho_{A}(r_1,r'_1)_{l=2}
\sim
R_{1p_{3/2}}(r_1)
R_{1p_{3/2}}(r'_1)
\frac{ e^{-B(r_1-r'_1)^2/2}}{r_1 r'_1}
\rho_0(r_m)
\int_{-\infty}^{\infty} dr'_2\, 
e^{-2Br'_2{}^2}.
\end{equation}
Finally, introducing the asymptotic behavior of the radial functions
\begin{eqnarray}
R_{1p_{3/2}}(r_1)
&\sim&
 \frac{e^{-kr_1}}{r_1}
\\
\rho(r_m)
&\sim&
 \frac{e^{-2kr_m}}{r_m^2}
= 4 \frac{e^{-k(r_1+r'_1)}}{(r_1+r'_1)^2},
\end{eqnarray}
we obtain the following asymptotic expression 
\begin{equation}
\rho_{A}(r_1,r'_1)_{l=2}
\sim
\frac{ \exp{(-2kr_1)}\exp{(-2kr'_1)}
       \exp\left( \frac{-B (r_1-r'_1)^2 }{2} \right)
     }
     { r_1^2 r'_1{}^2 (r_1+r'_1)^2   
     }
\kern 1cm 
r_1,r'_1 \rightarrow\infty.
\end{equation}
A similar expression can be obtained for diagram B of fig.~2.


\end{document}